\documentclass[fleqn,usenatbib]{mnras}

\usepackage{newtxtext}
\usepackage{amssymb}
\usepackage{newtxmath}
\usepackage{booktabs}
\usepackage{pdflscape}
\usepackage{pifont}
\usepackage{enumitem}
\usepackage{subfigure}
\usepackage{subcaption}
\usepackage{xcolor}

\usepackage[T1]{fontenc}

\DeclareRobustCommand{\VAN}[3]{#2}
\let\VANthebibliography\thebibliography
\def\thebibliography{\DeclareRobustCommand{\VAN}[3]{##3}\VANthebibliography}

\usepackage{graphicx}	
\usepackage{amsmath}	

\title[New radio rings discovered with MeerKAT]{MeerKAT discovery of 164 compact radio rings toward the Galactic Plane}

\author[C. Bordiu et al.]{C. Bordiu,$^{1}$\thanks{E-mail: cristobal.bordiu@inaf.it}
F. Bufano,$^{1}$
G. Umana,$^{1}$
J. R. Rizzo,$^{2}$
C. Spingola,$^{3}$
C. Trigilio,$^{1}$
S. Loru,$^{1}$
M. D. Filipovic,$^{4}$
\newauthor
C. Buemi,$^{1}$
F. Cavallaro,$^{1}$
L. Cerrigone,$^{5}$
P. Leto,$^{1}$
A. Ingallinera,$^{1}$
S. Riggi,$^{1}$
A. C. Ruggeri,$^{1}$
\newauthor
Z. Smeaton,$^{4}$
P. A. Woudt$^{6}$
\\
$^{1}$ INAF $-$ Osservatorio Astrofisico di Catania, Via Santa Sofía 78, I$-$95123 Catania, Italy\\
$^{2}$ ISDEFE, Beatriz de Bobadilla 3, 28040 Madrid, Spain. \\
$^{3}$ INAF $-$ Istituto di Radioastronomia, Via Gobetti 101, I$-$40129, Bologna, Italy \\
$^{4}$ Western Sydney University, Locked Bag 1797, Penrith South DC, NSW 2751, Australia \\
$^{5}$ Joint ALMA Observatory, Alonso de Córdova 3107, Vitacura, Santiago 7630355, Chile \\
$^{6}$ Department of Astronomy, University of Cape Town, Private Bag X3, Rondebosch, Cape Town, 7701, South Africa
}

\date{Accepted XXX. Received YYY; in original form ZZZ}

\pubyear{2015}

\begin{document}
\label{firstpage}
\pagerange{\pageref{firstpage}--\pageref{lastpage}}
\maketitle

\begin{abstract}
We report the discovery of 164 compact (radius < 1 arcmin) radio rings using MeerKAT 1.3 GHz data from the SARAO MeerKAT Galactic Plane Survey ($l=2\degr-60\degr, 252\degr-358\degr$, $|b|\le1\fdg5$)\ and the Galactic Centre mosaic, from a search aimed at identifying previously uncatalogued radio sources. Within this sample, approximately 19 per cent of the rings contain a central point radio source. A multiwavelength analysis reveals a striking diversity: about 40 per cent  of the rings enclose an isolated infrared point source, 50 per cent exhibit an extended counterpart in the mid- or far-infrared, and several are only detected in the radio band. We found that 17 per cent of the rings in the sample are positionally coincident (within 5 arcsec) with known entries in SIMBAD, including unclassified infrared sources, spiral galaxies, young stellar objects and long-period variable candidates. Based on these matches and exploiting ancillary multiwavelength data and catalogues, we explore several formation scenarios for the rings, such as  H\textsc{ii} regions, planetary nebulae, mass-loss relics from evolved massive stars, supernova remnants, nova shells, galaxies, galaxy cluster lenses and odd radio circles. Tentative classifications are proposed for nearly 60 per cent of the sample. These results highlight the potential of MeerKAT to uncover previously undetected compact radio structures and, particularly, recover missing Galactic radio-emitting objects.
\end{abstract}




\section{Introduction}

Circumstellar shells are prominent manifestations of the complex interplay between evolved stars of all masses and the interstellar medium (ISM). These conspicuous structures are mass loss relics, formed when fast stellar winds sweep up pre-existing circumstellar material or in sporadic outbursts that expel large amounts of matter over short timescales \citep{chu1991}. Owing to their dust and ionised gas content, circumstellar shells can be detected at infrared \citep{wac2010} and radio \citep{dun02} wavelengths, appearing as rings of emission surrounding their central sources---sometimes completely detached. The study of shell morphologies, kinematics and composition from a multiwavelength perspective is essential for reconstructing the mass-loss history of the progenitor stars. This, in turn, places tighter constraints on their evolutionary pathways \citep{riz2008,uma10,uma2011,bor2021}.

Over the past few decades, the blind search for ring-shaped structures in wide area infrared surveys \citep{chu06,miz2010} has proven to be one of the most effective methods for identifying circumstellar shells, and therefore, new evolved star candidates. This is exemplified by the prolific results of the \textit{Spitzer} MIPSGAL 24 $\mu$m survey of the Galactic Plane, in which nearly 400 ring- and disk-like structures (or \lq bubbles\rq) were detected. While many were confirmed as planetary nebulae (PNe)\citep{miz2010}, spectroscopic follow-ups of their associated central sources led to the identification of a subset as circumstellar shells around evolved massive stars, like Luminous Blue Variables (LBV) and Wolf-Rayet (WR) stars \citep{gva2010,wac2010,fla11,str2012,fla14}.

Ring-like structures are also commonly found in deep radio continuum surveys of the Galactic Plane. The largest sources usually correspond to either H\textsc{ii} regions or supernova remnants (SNRs) ---with sizes ranging from several arcmin to a few degrees---. In contrast, compact rings ($\lesssim$1 arcmin) may trace ionised ejecta around evolved stars, in the form of PNe or LBV/WR nebulae \citep{ing2016, ing2019}. So far, though, the potential of radio surveys to blindly search for ring sources has been somewhat limited by the inherent tradeoff between angular resolution, sensitivity, and sky coverage, preventing the faintest or more compact sources from being detected or resolved. 

These limitations are now being rapidly overcome with the advent of SKA precursors. Instruments such as  ASKAP \citep{joh08} and MeerKAT \citep{jon09, jon16} represent a significant leap in terms of imaging fidelity and survey speed, charting extensive areas of the sky with angular resolutions almost comparable to those of optical and infrared surveys, and sensitivities of the order of a few 10s $\mu$Jy beam$^{-1}$. In addition to their obvious value for a more accurate characterization of known evolved stars, these instruments show a great potential for unexpected discoveries, as evidenced by the detection of the so-called \lq Odd Radio Circles\rq\, (ORCs, \citealt{nor21a,nor21b,kor21}) and other related ring-like objects of unclear astrophysical nature (e.g., the Kýklos ring, \citealt{bor24}).

Wide-area surveys conducted with ASKAP and MeerKAT \citep{hop25,goe23} may therefore provide access to unobserved populations of evolved stars, especially those hidden in dust-obscured regions of the Galactic Plane, beyond the reach of optical or infrared instruments. The MeerKAT Galactic Centre mosaic \citep{hey2022}, a deep L-band (856--1712 MHz) survey of the central region of the Milky Way, offered a glimpse of this potential, revealing numerous faint and previously unseen \lq\lq low-angular diameter shells\rq\rq. Motivated by this finding, we undertook a systematic search for these and other similar sources across the Galactic Plane, combining the Galactic Centre mosaic data with the SARAO MeerKAT Galactic Plane survey (SMGPS). The SMGPS is the deepest L-band (856--1712 MHz) continuum survey of the Milky Way to date \citep{goe23}, mapping almost half of the Galactic Plane ($l=2\degr-60\degr, 252\degr-358\degr$, $|b|\le1\fdg5$)\footnote{The survey is designed to follow the Galactic warp, involving a southward offset of $0\fdg5$ for some fields, effectively extending below $b=-1.5\degr$} \citep{goe23} with an angular resolution of $\sim$8 arcsec and a point source sensitivity of $\sim$20--30 $\mu$Jy beam$^{-1}$. The surveyed field is populated by nearly $\sim17000$ extended radio sources, of which $24$ per cent have been confidently associated with known Galactic objects---H\textsc{ii} regions, SNRs, LBV and WR nebulae, and PNe--- and $33$ per cent are most likely extragalactic \citep{bor25}. The remaining $43$ per cent, comprising $\sim$7000 radio sources, remain unclassified. This unexplored sample holds strong potential for the discovery of new Galactic sources, possibly including a significant number of evolved star candidates.

In this paper, we present the results of a search for low-angular diameter radio rings in MeerKAT data from the Galactic Centre mosaic and the SMGPS. We identify a total of 164 ring-like sources and provide their flux densities, sizes, and coordinates. The paper is organised as follows: Sect. \ref{sec:sample} describes the sample, selection criteria, flux density measurement methods, and morphological classification, including a discussion on spatial distribution and potential selection biases. Sect. \ref{sec:crossmatch} presents a multiwavelength analysis of the rings and explores possible physical scenarios for their origin. Finally, Sect. \ref{sec:conclusions} summarises our findings and outlines future steps.

\section{SAMPLE OF RADIO RINGS}
\label{sec:sample}

\subsection{Source selection procedure}
\label{sec:sel-criteria}

We examined the Galactic Centre mosaic and the 56 tiles of the SMGPS to identify previously unclassified low-angular diameter radio rings. In particular, we looked for sources meeting the following criteria:

\begin{enumerate}
    \item The source must display a ring morphology, regardless of whether a central object is present. This requires a certain degree of limb brightening, even if the brightness distribution is clumpy or asymmetric. This criterion excludes filled disks or uniformly bright (\lq flat\rq) bubbles. Moderate deviations from spherical symmetry are acceptable, but extreme aspect ratios are not.
    \item The source angular radius (or semimajor axis in the case of slightly elliptical sources) must not exceed 1 arcmin. The reason is twofold. First, the angular radius distribution of known evolved stars in the SMGPS extended source catalogue \citep{bor25} supports this cut: 90 per cent of PNe and 86 per cent of LBVs fall below the 1 arcmin threshold. Second, it minimizes overlap with SMGPS SNR candidates \citep{and25}, which mostly exceed this size.
    \item The source must lie in a region of the sky preferably free of major imaging artefacts such as strong sidelobe features around bright sources, or deep negative bowls caused by missing flux from extended emission (see \citealt{goe23} for details). Likewise, as much as possible, the sources should be sufficiently isolated from \lq contaminating\rq\, emission associated with nearby large-scale structures such as filaments or SNRs.
\end{enumerate}

The extraction procedure involved a systematic visual inspection of each survey tile using \texttt{CARTA}. To isolate new sources, known objects were excluded by overlaying region files from the following catalogues of Galactic sources: the MIPSGAL catalogue of Galactic bubbles \citep{miz2010}, the 2018 census of Luminous Blue Variables in the Local Group \citep{ric2018}, the Galactic Wolf-Rayet Star Catalogue\footnote{\url{http://pacrowther.staff.shef.ac.uk/WRcat/index.php}} \citep{ros2015},  the Hong Kong/AAO/Strasbourg H-alpha (HASH) Planetary Nebula Database\footnote{\url{http://202.189.117.101:8999/gpne/dbMainPage.php}} \citep{par2016}; the WISE catalogue of H\textsc{ii} regions\footnote{\url{https://astro.phys.wvu.edu/wise/}}\citep{and14}; and catalogues of Galactic SNRs \citep{gre19} and candidates from THOR \citep{and17}, GLEAM \citep{hur19}, GLOSTAR \citep{dok21}, EMU \citep{bal23} and SMGPS \citep{and25}. Any discrete ring of emission not coinciding with a known region and meeting the criteria described above was identified and added to the sample. The position and angular size of each ring were carefully determined by eye, fitting a circle or ellipse to the observed emission, with a typical precision estimated to be on the order of one-third of the beam size.

\begin{figure*}
    \centering
    \includegraphics[width=0.95\textwidth]{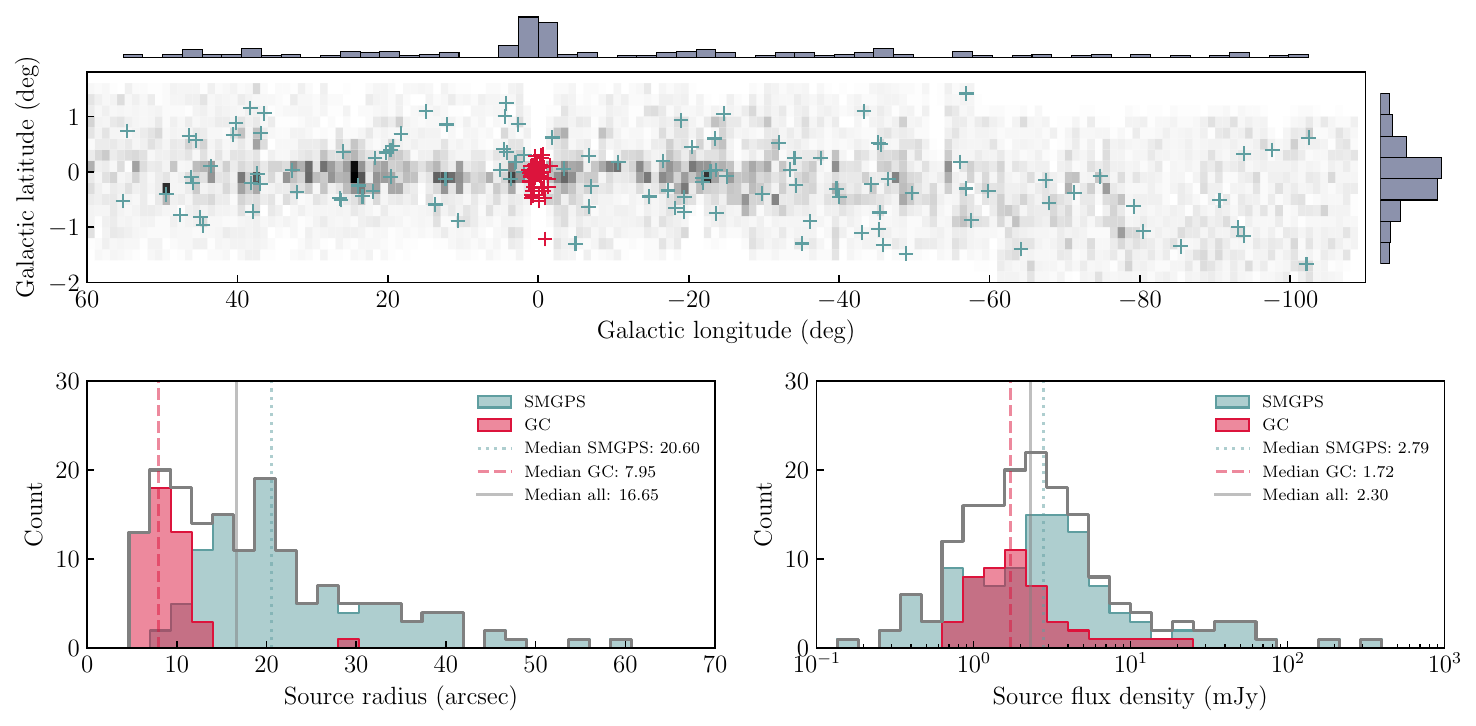}
    \caption{Top: sky distribution of the radio rings (SMGPS -- blue crosses, GC -- red crosses), overlaid on a 2D histogram representing the spatial density of extended sources in the SMGPS (colour scale), based on the catalogue by \protect\cite{bor25}. The histograms on top and right represent the marginal distribution of rings in $l$ and $b$. Bottom: histograms showing the distributions of ring angular radii (left) and flux densities (right) in the sample (cumulative histogram shown in grey). Vertical lines indicate the median values of each group.}
    \label{fig:sky-distribution}
\end{figure*}

A total of 164 radio rings were identified using this method: 116 in the SMGPS and 48 in the Galactic Centre mosaic. Table \ref{tab:ring-list} lists their assigned names (following IAU recommendation), angular sizes and flux densities. The full machine-readable table, including the rings' centroids Galactic coordinates, is available as online supplementary material.  Cutout images of all the rings are presented in Appendix \ref{app:images}, and their spatial distribution is shown in Fig. \ref{fig:sky-distribution}.

It is important to state that the primary goal of this work is the discovery and initial characterisation of a sample of new radio rings, rather than the creation of a statistically complete sample of ring-like sources in the surveyed area. Well-known objects hosting radio rings, such as some Luminous Blue Variables or Wolf-Rayet stars, have been intentionally excluded. Likewise, we make no claim regarding the detection completeness of the sample. The prevalence of extended background emission across the observed fields, often featuring multiple overlapping and intertwined structures, makes it likely that some rings were missed. Similarly, ring-shaped features appearing in complex regions, potentially resulting from pareidolic effects (e.g., substructure from SNRs) were rejected, even if some may correspond to real sources.

\subsection{Morphological classification}
\label{subsec:morpho}

\begin{figure*} 
    \centering
    \begin{minipage}[t]{0.3\textwidth}
        \centering
        \includegraphics[scale=0.28]{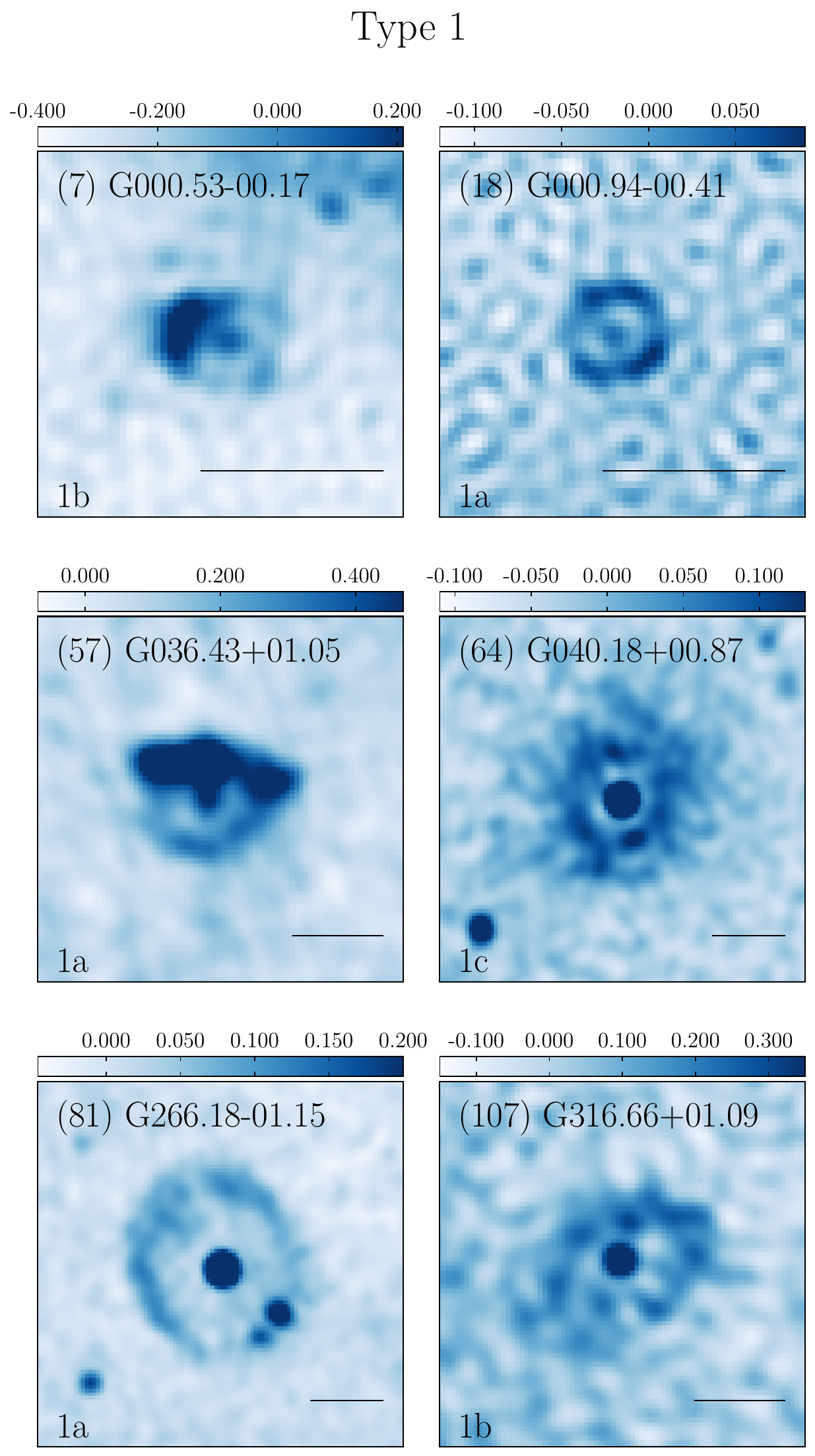} 
    \end{minipage}
    \hspace{2em}
    \begin{minipage}[t]{0.3\textwidth}
        \centering
        \includegraphics[scale=0.28]{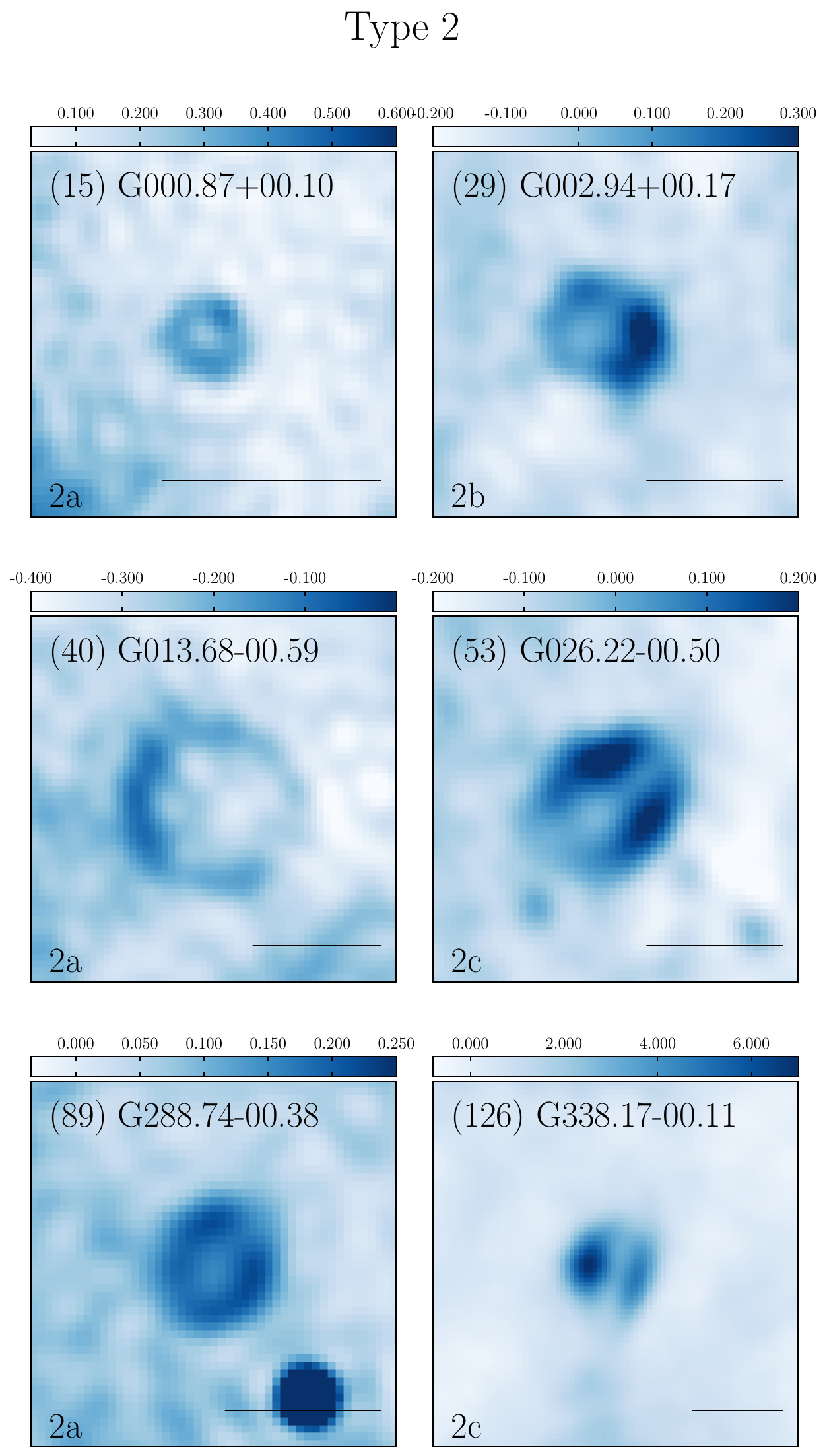} 
    \end{minipage}
    \hspace{2em}
    \begin{minipage}[t]{0.3\textwidth}
        \centering
        \includegraphics[scale=0.28]{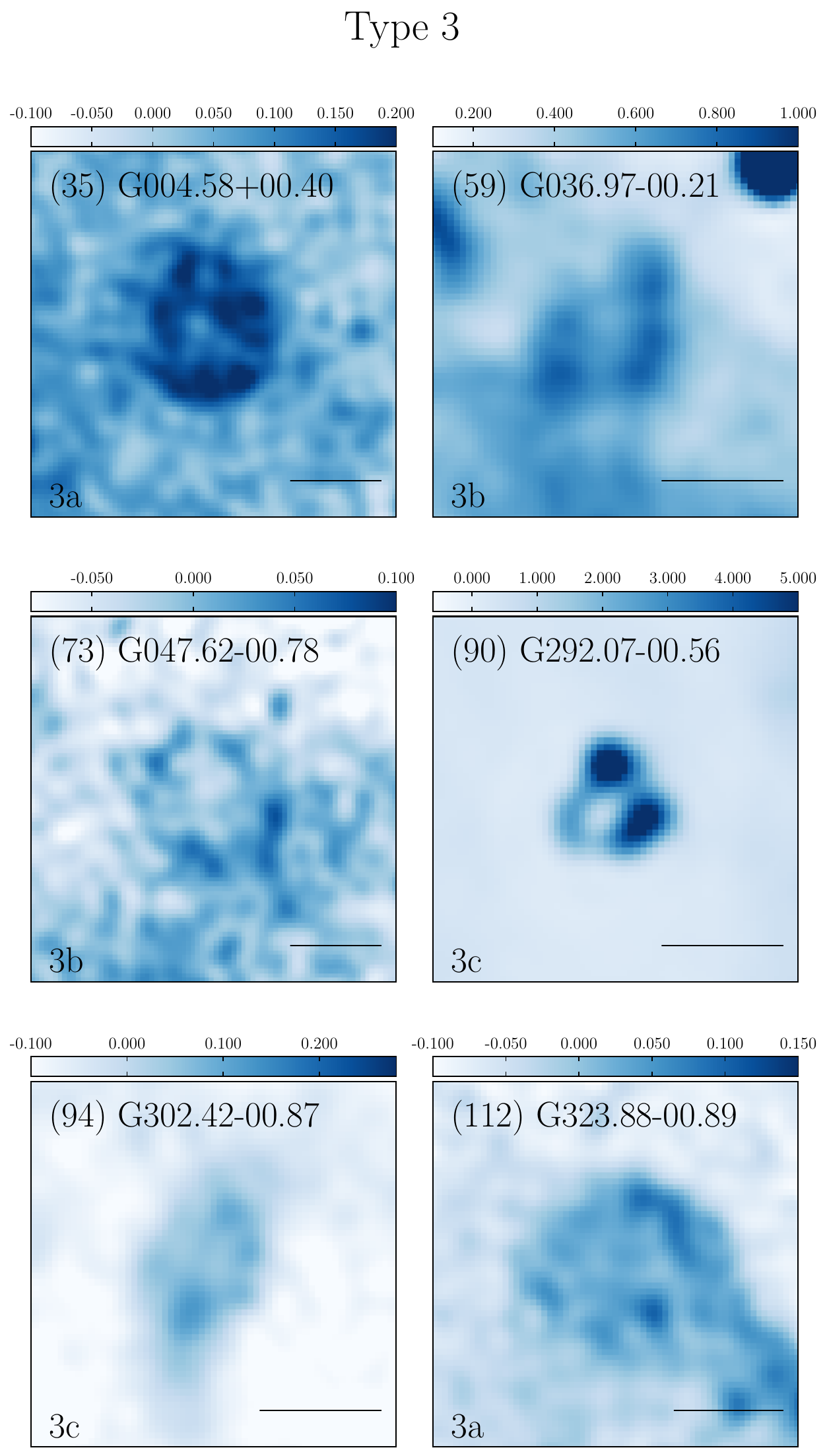} 
    \end{minipage}

    \caption{Examples of radio rings of type 1 (left), 2 (middle) and 3 (right). The name of each ring is shown in the top-left corner of each plot, with the morphological type indicated in the bottom-left. The scale bar represents 30 arcsec. Colour bar indicates surface brightness in units of mJy beam$^{-1}$.}
    \label{fig:examples-shells}
\end{figure*}

The morphology of the rings can offer insights into their nature. We conducted a visual examination of all extracted rings and grouped them based on shared features, establishing a preliminary taxonomy that loosely follows the classification scheme defined by \cite{miz2010} for the MIPSGAL rings, and forms a basis for future studies. In particular, we identified three main groups, as illustrated in Fig. \ref{fig:examples-shells}:

\begin{itemize}
\item \textit{Rings with central sources} (type 1, $\sim$19 per cent of the total) -- these are ring structures with a distinguishable point-like source near their centre. This group is further divided into \textit{regular rings} (type 1a, $\sim$29 per cent of type 1 rings), which are smooth and lack prominent features; \textit{irregular rings} (type 1b, $\sim$52 per cent), which include partial or broken rings as well as rings with asymmetrical brightness distributions; and \textit{diffuse rings}, faint rings without clear edges (type 1c, $\sim$19 per cent).
\item \textit{Rings without central sources} (type 2, $\sim$56 per cent of the total) -- these are rings lacking a central point source, including both \lq thin\rq\, and \lq thick\rq\, rings (nearly filled disks with a cavity toward the centre). This category is subdivided into: \textit{regular rings} (type 2a, $\sim$26 per cent of type 2 rings), \textit{irregular rings} (type 2b, $\sim$55 per cent), including broken rings or rings with uneven or asymmetric brightness distributions; and \textit{bilateral rings} (type 2c, $\sim$19 per cent), with symmetrically enhanced brightness distributions.
\item \textit{Miscellaneous rings} (type 3, $\sim$25 per cent of the total) -- this category includes rings that do not fit into the other classifications, such as \textit{diffuse rings} (type 3a, $\sim$34 per cent of type 3 rings), \textit{clumpy rings} (type 3b, $\sim$24 per cent), and \textit{peculiar rings} (type 3c, $\sim$42 per cent). \end{itemize}

Table \ref{tab:ring-breakdown} shows the source breakdown by type and subtype. It is important to note that this classification scheme is purely observational and may be significantly biased by the limited angular resolution, the intrinsic size of the sources and the distance uncertainty. For instance, compact clumpy rings may appear less clumpy than they actually are. Similarly, the ability to identify central point-like sources depends on the rings' angular size: in more extended rings, a central source is easier to resolve, whereas in compact rings, the central emission may blend with that of the ring itself, making it difficult to disentangle the two components. More critically, the sample may even include objects that are not truly ring-like, but merely appear so. For instance, extremely bent radio galaxies or unrelated clumps arranged in a roughly circular pattern can mimic ring-like structures in the MeerKAT images. This limitation is inherent to the data and cannot be mitigated without higher resolution observations. Its impact is likely more significant in the SMGPS, as the Galactic Centre mosaic data was CLEANed with a Briggs robust factor of $-1.5$ to increase resolution and suppress sidelobes \citep{hey2022}, resulting in an effective resolution of $4$ arcsec---compared to the 8 arcsec synthesised beam of SMGPS. 

\subsection{Flux density measurement}

To compute ring flux densities, we employed standard aperture photometry in \texttt{CARTA}\footnote{\url{https://cartavis.org/}}\footnote{ \url{https://zenodo.org/records/4905459}} (v4.0.0). In the general case, we defined a polygon region enclosing the emission from the ring, and computed the flux density $S$ by scaling the background subtracted brightness $B$:

\begin{equation}
    S = \frac{B}{\Omega}
\end{equation}

\noindent in which $\Omega$ is the scaling factor that relates the beam and pixel areas. This factor differs between SMGPS data and the Galactic Centre mosaic due to the different angular resolution \citep{hey2022}.

The background-subtracted source brightness is:

\begin{equation}
    B = \sum_{i=1}^{N_\mathrm{src}}B_\mathrm{src,i} - \sum_{i=1}^{N_\mathrm{src}}B_\mathrm{bkg}
\end{equation}

\noindent where $N_\mathrm{src}$ is the size of the source in pixels, $B_\mathrm{src,i}$ is the brightness of the $i$-th source pixel, and $B_\mathrm{bkg}$ is the local background estimate, set to the average brightness value in an annulus region of width 5 pixels that surrounds the source. A proper estimation of the local background is critical to obtain reliable source flux densities, especially due to the lack of zero-spacing information. This implies that bright extended sources (relatively ubiquitous in the Galactic Plane) are not properly sampled by the $uv-$coverage and end up surrounded by negative bowls. This is even more evident in SMGPS data due to the shallowness of the applied CLEAN \citep{goe23}. As a consequence, many rings have slightly negative local backgrounds (of the order of a few 10$^{-5}$ Jy beam$^{-1}$). In those rings with a central point-like source---provided it can be properly isolated from the ring structure---, we first performed a 2D Gaussian fitting to the central object. Then, we applied aperture photometry as described above on the residual image, i.e, the image with the central source subtracted. The flux density uncertainty was computed by error propagation, as the quadrature sum of the flux calibration uncertainty ($\sim$5 per cent, \citealt{goe23}) and the statistical uncertainty derived from the background $rms$ noise, as measured in the annulus. Fig. \ref{fig:sky-distribution} (bottom panels) shows the histograms of angular radius and flux density for the rings, with median values of $\sim$17 arcsec and $\sim$2 mJy respectively.

\subsection{Spatial distribution and selection bias}

Fig. \ref{fig:sky-distribution} shows a clear overdensity of radio rings in the Galactic Centre. The SMGPS spans approximately 500 deg$^2$ and contains 116 radio rings, corresponding to a density of $\sim$0.23 deg$^{-2}$. In contrast, the Galactic Centre mosaic covers only 6.5 deg$^{2}$, but contains 48 rings, resulting in a much higher ring density of 7.38 deg$^{-2}$. This pronounced excess may reflect a genuine overabundance of ring sources toward the Galactic Centre. If most rings are associated with different stages of stellar evolution, a higher detection rate in this region is expected, simply due to the higher stellar density. Additionally, the line-of-sight path length through the Milky Way is longer toward the Galactic Centre, increasing the chance of finding sources. However, observational biases could also contribute: the Galactic Centre mosaic has twice the angular resolution and better sensitivity than the SMGPS, allowing for detecting fainter or more compact rings.

The histograms in Fig. \ref{fig:sky-distribution}, show that the flux density distributions of the two populations are broadly similar. Even if the Galactic Centre mosaic is more sensitive,  most sources in both samples fall in the 1--10 mJy range, and the SMGPS distribution shows a more extended tail toward lower flux densities. This suggests that sensitivity is not the dominant factor driving the observed differences. Instead, a strong angular resolution bias is evident. The Galactic Centre mosaic contains many rings with much smaller angular radii (median $\sim$8 arcsec), which would likely appear unresolved or blended at the resolution of the SMGPS (where the median radius is $\sim$21 arcsec).

\begin{landscape}
\begin{table}
\caption{List of detected radio rings ordered by increasing Galactic longitude. Rings 1--25 and 142-164 extracted from the Galactic Centre mosaic. For each ring, the table lists its unique ID, IAU designation, approximate radius, flux densities at 1.3 GHz and associated uncertainties for the ring and central point source (if present), morphology type, and multiwavelength crossmatch information. Morphology types, defined in Sect. \ref{subsec:morpho}, are 1--ring with central radio source (a=regular, b=irregular, c=diffuse), 2--ring without central radio source (a=regular, b=irregular, c=bipolar) and 3--miscellaneous ring (a=diffuse, b=clumpy, c=peculiar). In the 2MASS and GLIMPSE columns, a \checkmark denotes the presence of a non-confused point source in the corresponding catalogue, $\sim$ indicates a confused source due to crowding, and \ding{53} indicates no match. In the 8/24/70 $\mu$m columns, a \checkmark indicates unambiguously detected extended emission, $?$ denotes uncertain association, and \ding{53} indicates non-detection. OUT indicates that the source lies outside the coverage area of the corresponding survey. The last two columns indicate (i) whether a match was found in SIMBAD within a 5 arcsec radius, and (ii) a proposed classification tag based on the discussion in Sect. \ref{sec:crossmatch}. See main text for further details, and Appendix \ref{app:notes} for notes on individual sources.}
\label{tab:ring-list}
\footnotesize

\begin{tabular}{llrclllc@{\hspace{5pt}}cc@{\hspace{5pt}}c@{\hspace{5pt}}cll}
\toprule
 ID & GName & Radius & PS & $S_\mathrm{ring}$ & $S_\mathrm{PS}$ & Type & \multicolumn{2}{|c|}{Point source} & \multicolumn{3}{|c|}{Extended emission} & SIMBAD & Classification\\
\cline{8-10} \cline{11-12}
\\[-0.4ex]
 &       & (arcsec) &  & (mJy) & (mJy) & & 2MASS & GLIMPSE & 8 $\mu$m &  24 $\mu$m & 70 $\mu$m &  \\
\midrule
     1 & G000.21+00.12 &    13.1 &   N &   2.94 (0.3) &          -- &    2a &  \ding{53} &       $\sim$ &  \ding{53} & \checkmark &  \ding{53} &                                       &                                \\
     2 & G000.44+00.28 &    12.1 &   Y &  1.32 (0.12) & 0.07 (0.02) &    1a & \checkmark &       $\sim$ &  \ding{53} & \checkmark &  \ding{53} &                                       &                                \\
     3 & G000.45$-$00.00 &     5.7 &   N &  9.26 (0.54) &          -- &    2b & \checkmark &       $\sim$ &  \ding{53} & \checkmark &  \ding{53} &              [FPF2021] 44  5 1 (Star) &                   PN/Massive star \\
     4 & G000.50+00.12 &     7.7 &   N &  1.85 (0.16) &          -- &    3c &       $\sim$ &       $\sim$ &  \ding{53} &          ? &  \ding{53} &                                       &                                \\
     5 & G000.50$-$00.11 &     7.2 &   N &   2.71 (0.2) &          -- &    2b &  \ding{53} &       $\sim$ &  \ding{53} &          ? &  \ding{53} &                                       &                  PN              \\
     6 & G000.52+00.04 &     7.6 &   N &   1.9 (0.21) &          -- &    2c &       $\sim$ &       $\sim$ &  \ding{53} &  \ding{53} &  \ding{53} &                                       &            PN                    \\
     7 & G000.53$-$00.17 &    11.5 &   Y &  4.04 (0.32) &  0.4 (0.08) &    1b &       $\sim$ &       $\sim$ &  \ding{53} & \checkmark & \checkmark &      ISOGAL-P J174733.0$-$283411 (YSOc) &                  H\textsc{ii} region \\
     8 & G000.58$-$00.16 &     7.7 &   N &  6.76 (0.38) &          -- &    2b &       $\sim$ &       $\sim$ &  \ding{53} & \checkmark & \checkmark &                   SSTGC 830399 (YSOc) &                  H\textsc{ii} region \\
     9 & G000.61$-$00.26 &    10.3 &   Y &   0.65 (0.1) & 0.15 (0.03) &    1a & \checkmark &       $\sim$ &  \ding{53} &  \ding{53} & \checkmark &                                       &                                \\
    10 & G000.61+00.01 &     5.1 &   N &  1.26 (0.11) &          -- &    2c & \checkmark & \checkmark &  \ding{53} & \checkmark &  \ding{53} &                                       &                       PN         \\
    11 & G000.70$-$00.01 &    10.9 &   N &  1.95 (0.21) &          -- &    3c &       $\sim$ &       $\sim$ &  \ding{53} &  \ding{53} &  \ding{53} &                                       &                                \\
    12 & G000.74$-$00.43 &     9.7 &   Y &   0.98 (0.13) &  0.22 (0.02) &    1a & \checkmark &       $\sim$ &  \ding{53} & \checkmark &          ? &                                       &                   Massive star \\
    13 & G000.81$-$00.06 &     6.7 &   N &  1.17 (0.11) &          -- &    2b &       $\sim$ &       $\sim$ &  \ding{53} &  \ding{53} &          ? &                                       &                   PN             \\
    14 & G000.82$-$00.34 &     6.9 &   N &  1.69 (0.11) &          -- &    2b &  \ding{53} &       $\sim$ &  \ding{53} & \checkmark &  \ding{53} &                                       &                  PN              \\
    15 & G000.87+00.10 &     7.9 &   N &  1.63 (0.15) &          -- &    2a & \checkmark &       $\sim$ &  \ding{53} & \checkmark &  \ding{53} &                                       &                   Massive star \\
    16 & G000.89$-$00.06 &     8.2 &   N &  2.19 (0.15) &          -- &    2a & \checkmark &       $\sim$ &  \ding{53} &  \ding{53} &  \ding{53} &                                       &                                \\
    17 & G000.92$-$00.46 &    10.8 &   Y? &  1.04 (0.11) &          -- &    1c &       $\sim$ &       $\sim$ &  \ding{53} & \checkmark &  \ding{53} &                                       &                   Massive star \\
    18 & G000.94$-$00.41 &     9.4 &   Y &  0.89 (0.09) & 0.15 (0.03) &    1a &       $\sim$ &       $\sim$ &  \ding{53} & \checkmark &          ? &                                       &                   Massive star \\
    19 & G001.02+00.10 &     7.0 &   N &  1.23 (0.11) &          -- &    3b &       $\sim$ &       $\sim$ &  \ding{53} &          ? &          ? &                                       &                                \\
    20 & G001.07+00.01 &     6.7 &   N &  1.56 (0.12) &          -- &    2b &  \ding{53} &       $\sim$ &  \ding{53} &  \ding{53} &  \ding{53} &                    SPICY 64996 (YSOc) &                  H\textsc{ii} region \\
    21 & G001.07$-$00.08 &     7.0 &   N &  1.91 (0.18) &          -- &    2a &  \ding{53} &       $\sim$ &  \ding{53} &  \ding{53} &  \ding{53} &                    SPICY 65507 (YSOc) &                  H\textsc{ii} region \\
    22 & G001.16$-$00.02 &     7.4 &   N &   2.1 (0.18) &          -- &    2a &  \ding{53} &       $\sim$ &  \ding{53} &  \ding{53} &  \ding{53} &                                       &                                \\
    23 & G001.17$-$00.16 &     8.3 &   N &   1.01 (0.1) &          -- &    2b &       $\sim$ &       $\sim$ &  \ding{53} &  \ding{53} &  \ding{53} &                                       &                       PN         \\
    24 & G001.22$-$00.18 &     7.0 &   N &  0.92 (0.08) &          -- &    2b &       $\sim$ &       $\sim$ &  \ding{53} &  \ding{53} &  \ding{53} &                                       &                        PN        \\
    25 & G001.26+00.02 &     6.7 &   N &  1.43 (0.14) &          -- &    2b &  \ding{53} &       $\sim$ &  \ding{53} & \checkmark &  \ding{53} &                                       &                      PN          \\
    26 & G001.85+00.30 &    14.3 &   N &  1.94 (0.27) &          -- &    2b &       $\sim$ &       $\sim$ &  \ding{53} & \checkmark &  \ding{53} &                                       &                        PN        \\
    27 & G001.92$-$00.01 &    11.2 &   N &  0.42 (0.12) &          -- &    2b &       $\sim$ &       $\sim$ &  \ding{53} &  \ding{53} &  \ding{53} &                                       &                         PN       \\
    28 & G002.65+00.85 &    14.3 &   N &  0.45 (0.05) &          -- &    3c &  \ding{53} &       $\sim$ &  \ding{53} &  \ding{53} &          ? &                                       &                                \\
    29 & G002.94+00.17 &    12.7 &   N &  1.98 (0.14) &          -- &    2b &       $\sim$ &       $\sim$ &  \ding{53} &  \ding{53} &  \ding{53} &                                       &                         PN       \\
    30 & G003.06+00.17 &    37.5 &   N &  5.51 (0.57) &          -- &    2b &       $\sim$ &       $\sim$ &  \ding{53} &  \ding{53} &  \ding{53} &                                       &            ORC                    \\
    31 & G003.66$-$00.13 &    14.1 &   N &  3.74 (0.36) &          -- &    2b &       $\sim$ &       $\sim$ &  \ding{53} & \checkmark & \checkmark &                                       &                             PN \\
    32 & G004.21+00.35 &    26.3 &   N &  2.34 (0.19) &          -- &    3a &  \ding{53} &  \ding{53} &  \ding{53} &  \ding{53} &  \ding{53} &                                       &                             PN \\
    33 & G004.25+01.23 &    18.3 &   N &   1.6 (0.11) &          -- &    2b &  \ding{53} &  \ding{53} &  \ding{53} & \checkmark &  \ding{53} &                                       &                                \\
    34 & G004.41+00.99 &    11.6 &   N &  0.76 (0.08) &          -- &    2b &  \ding{53} &       $\sim$ &  \ding{53} & \checkmark &  \ding{53} &                                       &                             PN \\
\midrule
\end{tabular}
\end{table}
\end{landscape}

\begin{landscape}
\begin{table}
\contcaption{}
\footnotesize
\setlength{\tabcolsep}{5pt}
\begin{tabular}{llrclllc@{\hspace{5pt}}cc@{\hspace{5pt}}c@{\hspace{5pt}}cll}
\toprule
 ID & GName & Radius & PS & $S_\mathrm{ring}$ & $S_\mathrm{PS}$ & Type & \multicolumn{2}{|c|}{Point source} & \multicolumn{3}{|c|}{Extended emission} & SIMBAD & Classification\\
\cline{8-10} \cline{11-12}
\\[-0.4ex]
 &       & (arcsec) &  & (mJy) & (mJy) & & 2MASS & GLIMPSE & 8 $\mu$m &  24 $\mu$m & 70 $\mu$m &  \\
\midrule
    35 & G004.58+00.40 &    26.3 &   N &  2.43 (0.18) &          -- &    3a &      $\sim$&      $\sim$&  \ding{53} &  \ding{53} &  \ding{53} &                                       &                             PN \\
    36 & G005.05+00.02 &    13.7 &   N &  4.54 (0.29) &          -- &    2b &  \ding{53} &      $\sim$&  \ding{53} & \checkmark & \checkmark &                                       &                             PN \\
    37 & G010.67$-$00.88 &    23.1 &   N &  3.72 (0.26) &          -- & 2b/2c &  \ding{53} &  \ding{53} &  \ding{53} & \checkmark &  \ding{53} &                                       &                             PN \\
    38 & G012.19+00.85 &    17.4 &   N &  3.78 (0.21) &          -- &    2a & \checkmark &      $\sim$&  \ding{53} &  \ding{53} &  \ding{53} &                                       &                             PN \\
    39 & G012.33$-$00.13 &    19.1 &   Y &  5.42 (0.42) & 0.56 (0.22) &    1b & \checkmark &      $\sim$&  \ding{53} &          ? & \checkmark &                                       &                   Massive star \\
    40 & G013.68$-$00.59 &    23.2 &   N &  1.38 (0.23) &          -- &    2a & \checkmark & \checkmark &  \ding{53} &  \ding{53} &  \ding{53} &                                       &           ORC                     \\
    41 & G014.93+01.09 &    18.6 &   Y? &  1.36 (0.11) &          -- &   1b? & \checkmark &  \ding{53} &  \ding{53} &  \ding{53} &  \ding{53} &                                       &                 \\
    42 & G018.25+00.69 &    18.7 &   N &  2.24 (0.19) &          -- &    2b & \checkmark & \checkmark &  \ding{53} & \checkmark &  \ding{53} &        2MASS J18214114$-$1242424 (LPVc) &              PN / Massive star \\
    43 & G019.34+00.46 &    25.6 &   N &  2.16 (0.23) &          -- &    3a & \checkmark &      $\sim$&  \ding{53} & \checkmark & \checkmark &                                       &                                \\
    44 & G019.60$-$00.08 &    26.6 &   N &  59.0 (3.38) &          -- &    2b &  \ding{53} & \checkmark &  \ding{53} & \checkmark & \checkmark &                                       &                     H\textsc{ii} region \\
    45 & G019.65+00.39 &    11.5 &   N &  1.32 (0.09) &          -- &    2a & \checkmark &      $\sim$&  \ding{53} &          ? &  \ding{53} &                                       &                                \\
    46 & G020.24+00.34 &    22.5 &   N &  2.69 (0.27) &          -- &    3c & \checkmark &      $\sim$&  \ding{53} &  \ding{53} &  \ding{53} &     2MASS J18264481$-$1106216 (Star) &                                \\
    47 & G021.72+00.24 &    34.6 &   N &  2.93 (0.24) &          -- &    3a & \checkmark &      $\sim$&  \ding{53} & \checkmark & \checkmark &                                       &                     H\textsc{ii} region \\
    48 & G021.97$-$00.35 &    31.8 &   N &  2.71 (0.34) &          -- &    3a &      $\sim$&      $\sim$&  \ding{53} &  \ding{53} & \checkmark &                                       &                                \\
    49 & G023.37$-$00.44 &    12.1 &   N &  1.86 (0.23) &          -- &   2b & \checkmark &  \ding{53} &  \ding{53} &  \ding{53} &  \ding{53} &                                       &                                \\
    50 & G023.92$-$00.24 &    35.4 &   N &   5.7 (0.62) &          -- &    2b &  \ding{53} &      $\sim$&  \ding{53} &  \ding{53} &  \ding{53} &                                       &                                \\
    51 & G023.93$-$00.26 &    20.9 &   N &  2.98 (0.34) &          -- &    3a &      $\sim$&      $\sim$&  \ding{53} &  \ding{53} &  \ding{53} &                                       &                                \\
    52 & G025.94+00.36 &    34.7 &   Y? &   5.29 (0.6) &   -- &    1b &      $\sim$&      $\sim$&  \ding{53} &  \ding{53} &          ? &                                       &                                \\
    53 & G026.22$-$00.50 &    15.8 &   N &  2.79 (0.49) &          -- &    2c & \checkmark & \checkmark &  \ding{53} & \checkmark &  \ding{53} &                                       &                             PN \\
    54 & G026.41$-$00.47 &    13.0 &   N &  0.34 (0.02) &          -- &    2c &      $\sim$&      $\sim$&  \ding{53} &  \ding{53} &          ? &                                       &                             PN   \\
    55 & G032.09$-$00.36 &    17.2 &   N &  0.53 (0.07) &          -- &    2b &  \ding{53} &  \ding{53} &  \ding{53} &          ? &  \ding{53} &                                       &                                \\
    56 & G032.71+00.02 &    32.1 &   N &  9.94 (0.87) &          -- &    2a &      $\sim$& \checkmark &  \ding{53} & \checkmark & \checkmark & [SPK2012] MWP1G032714+000291 (Bubble) &                H\textsc{ii} region / PN \\
    57 & G036.43+01.05 &    20.0 &   Y &  2.76 (0.19) & 0.27 (0.05) &    1a &      $\sim$& \checkmark & \checkmark &        OUT & \checkmark &        2MASS J18540458+0335446 (LPVc) &                   Massive star \\
    58 & G036.89+00.69 &    45.8 &   N &  5.41 (0.39) &          -- &    3c &  \ding{53} &  \ding{53} &          ? &          ? &          ? &                                       &                                \\
    59 & G036.97$-$00.21 &    19.7 &   N &  3.47 (0.42) &          -- &    3b & \checkmark &      $\sim$& \checkmark & \checkmark & \checkmark &                                       &                     H\textsc{ii} region \\
    60 & G037.38$-$00.03 &    14.0 &   N &  3.17 (0.65) &          -- &    2b &  \ding{53} & \checkmark &  \ding{53} &          ? &  \ding{53} &                                       &                                \\
    61 & G037.97$-$00.72 &    40.7 &  Y? &   2.99 (0.3) &          -- &   3a? & \checkmark & \checkmark &  \ding{53} &  \ding{53} &  \ding{53} &                                       &                                \\
    62 & G038.21$-$00.20 &    41.2 &   N &  7.87 (0.64) &          -- &    3a &      $\sim$& \checkmark &  \ding{53} &  \ding{53} &  \ding{53} &                                       &                                \\
    63 & G038.28+01.15 &    16.6 &   N &  2.35 (0.13) &          -- &    2b & \checkmark &  \ding{53} &        OUT &        OUT &        OUT &                                       &                                \\
    64 & G040.18+00.87 &    28.2 &   Y &  2.78 (0.22) & 1.26 (0.07) &    1c & \checkmark & \checkmark & \checkmark &  \ding{53} &  \ding{53} &                                       &  Galaxy                              \\
    65 & G040.56+00.66 &    33.4 &   N &  2.63 (0.28) &          -- &    2a &  \ding{53} &  \ding{53} &  \ding{53} &  \ding{53} &  \ding{53} &                                       &                                \\
    66 & G043.55+00.10 &    26.5 &   Y &  0.40 (0.16) & 0.06 (0.05) &    1c &  \ding{53} & \checkmark &  \ding{53} &  \ding{53} &  \ding{53} &                                       &                                \\
    67 & G044.59$-$00.96 &     8.3 &   N &  0.14 (0.06) &          -- &    2b & \checkmark & \checkmark &  \ding{53} & \checkmark &  \ding{53} &                                       &                                \\
    68 & G044.96$-$00.82 &    37.2 &   N &  4.03 (0.26) &          -- &    2a & \checkmark & \checkmark &  \ding{53} &  \ding{53} &  \ding{53} &                                       &                                \\

    69 & G045.51+00.56 &    20.8 &  Y? &  0.62 (0.08) &          -- &   1b? &  \ding{53} &  \ding{53} &  \ding{53} &  \ding{53} &  \ding{53} &                                       &                                \\
    70 & G045.97$-$00.20 &    21.4 &   N &  1.78 (0.18) &          -- &    2b &  \ding{53} & \checkmark &  \ding{53} &  \ding{53} &  \ding{53} &                                       &                                \\
    71 & G046.16$-$00.09 &    17.1 &   N &  1.31 (0.13) &          -- &    2c &       $\sim$ & \checkmark & \checkmark & \checkmark & \checkmark &                                       &                                \\
    72 & G046.42+00.65 &    20.3 &   N &   3.68 (0.2) &          -- &    2c &  \ding{53} &  \ding{53} &  \ding{53} & \checkmark & \checkmark &            IRAS 19117+1212 (Infrared) &                             PN \\
    \midrule
\end{tabular}
\end{table}
\end{landscape}

\begin{landscape}
\begin{table}
\contcaption{}
\footnotesize
\setlength{\tabcolsep}{5pt}
\begin{tabular}{llrclllc@{\hspace{5pt}}cc@{\hspace{5pt}}c@{\hspace{5pt}}cll}
\toprule
 ID & GName & Radius & PS & $S_\mathrm{ring}$ & $S_\mathrm{PS}$ & Type & \multicolumn{2}{|c|}{Point source} & \multicolumn{3}{|c|}{Extended emission} & SIMBAD & Classification\\
\cline{8-10} \cline{11-12}
\\[-0.4ex]
 &       & (arcsec) & & (mJy) & (mJy) & & 2MASS & GLIMPSE & 8 $\mu$m &  24 $\mu$m & 70 $\mu$m &  \\
\midrule
73 & G047.62$-$00.78 &    33.5 &   N &  0.88 (0.15) &          -- &    3b & \checkmark & \checkmark &  \ding{53} &  \ding{53} &  \ding{53} &                                       &                                \\
    74 & G049.46$-$00.40 &    21.9 &   N &  73.2 (4.89) &          -- &    2b & \checkmark & \checkmark &  \ding{53} &  \ding{53} &        OUT &                                       &                                \\
    75 & G054.63+00.73 &    47.4 &   Y &  5.73 (0.36) & 0.19 (0.04) &    1c &      $\sim$& \checkmark &  \ding{53} &  \ding{53} &  \ding{53} &                                       &                                \\
    76 & G055.21$-$00.52 &    26.7 &  Y? &  0.87 (0.08) & 0.03 (0.03) &   1c? & \checkmark &  \ding{53} & \checkmark & \checkmark & \checkmark &                                       &                                \\
    77 & G257.56+00.61 &    19.5 &   Y &  0.49 (0.05) &          -- &    1b &  \ding{53} &  \ding{53} &  \ding{53} &        OUT &        OUT &                                       &                                \\
    78 & G257.84$-$01.66 &    32.2 &   N &  1.16 (0.09) &          -- &    2b &  \ding{53} &  \ding{53} &        OUT &        OUT &  \ding{53} &                                       &                                \\
    79 & G262.39+00.39 &    41.6 &   Y &  20.1 (1.07) & 0.55 (0.11) &    1b & \checkmark &  \ding{53} & \checkmark &        OUT &        OUT &    ZOA J084521.622$-$421845.17 (Galaxy) &   Galaxy                             \\
    80 & G266.18+00.32 &    35.8 &   Y &  31.4 (1.72) & 2.59 (0.15) &    1b & \checkmark &  \ding{53} & \checkmark &        OUT &        OUT &    ZOA J085828.676$-$451630.99 (Galaxy) &  Galaxy                              \\
    81 & G266.18$-$01.15 &    39.7 &   Y &   4.04 (0.3) & 1.67 (0.09) &    1a & \checkmark &  \ding{53} &  \ding{53} &        OUT &  \ding{53} &        2MASS J08520086$-$4614178 (YSOc) &                  H\textsc{ii} region \\
    82 & G266.96$-$01.00 &    20.8 &   Y &   4.07 (0.3) & 0.45 (0.11) &    1a &  \ding{53} &  \ding{53} &  \ding{53} &        OUT &  \ding{53} &               PSR J0855$-$4644 (Pulsar) &        Pulsar Wind Nebula                        \\
    83 & G269.40$-$00.51 &    25.0 &   N &  2.79 (0.16) &          -- &    3c & \checkmark &  \ding{53} & \checkmark &        OUT &          ? &                 HIZOA J0907$-$48 (AGNc) &                             Galaxy \\
    84 & G274.59$-$01.34 &    22.9 &   Y &  0.89 (0.08) & 0.22 (0.03) &    1c & \checkmark &  \ding{53} & \checkmark &        OUT &          ? &                                       &                Galaxy                \\
    85 & G279.55$-$01.07 &    38.4 &   Y &  3.98 (0.21) & 0.04 (0.03) &    1b & \checkmark &  \ding{53} & \checkmark &        OUT &          ? &                                       &                                \\
    86 & G280.80$-$00.62 &    18.2 &   N &  0.34 (0.03) &          -- &    2b &  \ding{53} &  \ding{53} &  \ding{53} &        OUT &  \ding{53} &                                       &                  Galaxy              \\
    87 & G285.25$-$00.07 &    24.9 &   N & 359.0 (20.2) &          -- &    2c & \checkmark &  \ding{53} & \checkmark &        OUT & \checkmark &        2MASS J10312191$-$5803345 (LPVc) &                   Massive star \\
    88 & G285.29$-$00.08 &    26.7 &   N & 190.0 (9.62) &          -- &    2b &  \ding{53} &  \ding{53} & \checkmark &        OUT & \checkmark &                                       &                             PN \\
    89 & G288.74$-$00.38 &    12.0 &   N &  0.75 (0.06) &          -- &    2a &  \ding{53} &  \ding{53} &  \ding{53} &        OUT &  \ding{53} &                                       &                                \\
    90 & G292.07$-$00.56 &    15.8 &   N &  23.7 (1.23) &          -- &    3c &  \ding{53} &  \ding{53} &  \ding{53} &        OUT &  \ding{53} &                                       &                             PN \\
    91 & G292.51$-$00.15 &    16.0 &   N &  1.45 (0.12) &          -- &    2b & \checkmark &  \ding{53} &          ? &  \ding{53} & \checkmark &                                       &                   Massive star \\
    92 & G295.81$-$01.39 &    20.9 &   Y &  0.86 (0.08) & 0.12 (0.02) &    1b &  \ding{53} &  \ding{53} &        OUT &        OUT &  \ding{53} &                                       &                                \\
    93 & G300.16$-$00.34 &    13.9 &   N &  0.72 (0.05) &          -- &    2c & \checkmark &      $\sim$& \checkmark & \checkmark & \checkmark &                                       &                                \\
    94 & G302.42$-$00.87 &    20.6 &   N &  1.98 (0.15) &          -- &    3c & \checkmark &  \ding{53} &  \ding{53} &  \ding{53} &  \ding{53} &                                       &               Galaxy                 \\
    95 & G303.08+01.41 &    25.0 &   N &  4.04 (0.23) &          -- &    3c &  \ding{53} &  \ding{53} &        OUT &        OUT &        OUT &                                       &                             PN \\
    96 & G303.14$-$00.30 &    22.0 &   N &  0.76 (0.06) &          -- &    2b & \checkmark & \checkmark &  \ding{53} &  \ding{53} &  \ding{53} &                                       &               Galaxy                 \\
    97 & G303.87+00.17 &    14.3 &   N &  1.04 (0.06) &          -- &    3b &      $\sim$&      $\sim$&  \ding{53} &  \ding{53} &  \ding{53} &                                       &                    \\
    98 & G310.29$-$00.38 &    20.1 &   N &  0.73 (0.06) &          -- &    3b &  \ding{53} &  \ding{53} &  \ding{53} & \checkmark & \checkmark &                                       &                                \\
    99 & G311.08$-$01.47 &    12.1 &   N &  2.97 (0.26) &          -- &    2c & \checkmark &  \ding{53} &        OUT &        OUT &        OUT &                                       &                             Galaxy \\
   100 & G313.45$-$00.13 &    18.9 &   N &  4.07 (0.28) &          -- &    2a & \checkmark & \checkmark &          ? & \checkmark & \checkmark &                                       &                H\textsc{ii} region / PN \\
   101 & G314.11$-$01.32 &    19.2 &   N &  1.93 (0.13) &          -- &    2b &      $\sim$&  \ding{53} &        OUT &        OUT &  \ding{53} &                                       &                             PN \\
   102 & G314.45+00.49 &    14.5 &   N &  0.45 (0.04) &          -- &    3c &  \ding{53} & \checkmark &  \ding{53} &  \ding{53} &  \ding{53} &                                       &                      Galaxy          \\
   103 & G314.59$-$00.73 &    14.5 &   N &  0.83 (0.06) &          -- &    3b &  \ding{53} & \checkmark &  \ding{53} &  \ding{53} &  \ding{53} &                                       &                                \\
   104 & G314.74$-$01.03 &    10.6 &   N &  0.37 (0.03) &          -- &    2c &  \ding{53} &  \ding{53} &  \ding{53} &  \ding{53} &  \ding{53} &                                       &                         PN       \\
   105 & G314.76+00.52 &    21.4 &   N &  1.07 (0.08) &          -- &    2b &  \ding{53} &  \ding{53} &  \ding{53} & \checkmark & \checkmark &                                       &                                \\
   106 & G315.71$-$00.22 &    39.6 &   N &  7.23 (0.42) &          -- &    2a &  \ding{53} & \checkmark &  \ding{53} & \checkmark &          ? &                                       &              Galaxy                  \\
   107 & G316.66+01.09 &    31.9 &   Y &  4.38 (0.36) &   0.9 (0.1) &    1b &      $\sim$& \checkmark &        OUT &        OUT &  \ding{53} &                                       &                   Massive star \\
   108 & G316.97$-$01.10 &    12.4 &   N &  0.84 (0.05) &          -- &    2a &  \ding{53} &  \ding{53} &  \ding{53} &        OUT & \checkmark &                                       &                             PN \\
   109 & G319.96$-$00.45 &    12.4 &   N &  1.69 (0.34) &          -- &    2b & \checkmark & \checkmark & \checkmark & \checkmark & \checkmark &        2MASS J15084437$-$5842399 (YSOc) &   H\textsc{ii} region / Massive star \\
   110 & G320.33$-$00.30 &    16.5 &   N &  4.75 (0.36) &          -- &    3c &  \ding{53} & \checkmark &          ? &          ? &          ? &                    SPICY 25893 (YSOc) &                    \\
\midrule
\end{tabular}
\end{table}
\end{landscape}

\begin{landscape}
\begin{table}
\contcaption{}
\footnotesize
\setlength{\tabcolsep}{5pt}
\begin{tabular}{llrclllc@{\hspace{5pt}}cc@{\hspace{5pt}}c@{\hspace{5pt}}cll}
\toprule
 ID & GName & Radius & PS & $S_\mathrm{ring}$ & $S_\mathrm{PS}$ & Type & \multicolumn{2}{|c|}{Point source} & \multicolumn{3}{|c|}{Extended emission} & SIMBAD & Classification\\
\cline{8-10} \cline{11-12}
\\[-0.4ex]
 &       & (arcsec) & & (mJy) & (mJy) & & 2MASS & GLIMPSE & 8 $\mu$m &  24 $\mu$m & 70 $\mu$m &  \\
\midrule
111 & G322.44+00.24 &    20.6 &   N &  13.1 (0.74) &          -- &    2a & \checkmark & \checkmark &  \ding{53} & \checkmark & \checkmark &                IRAS 15179$-$5638 (LPVc) &              PN / Massive star \\
   112 & G323.88$-$00.89 &    22.9 &   N &  2.41 (0.18) &          -- &    3a & \checkmark & \checkmark &  \ding{53} & \checkmark &  \ding{53} &                                       &                   Massive star \\
   113 & G324.92$-$01.29 &     9.2 &   N &  0.35 (0.05) &          -- &    2a &  \ding{53} &  \ding{53} &        OUT &        OUT &  \ding{53} &                                       &                                \\
   114 & G325.71$-$00.24 &    16.7 &   Y? &   7.8 (0.41) &          -- &    3c & \checkmark & \checkmark &  \ding{53} &          ? & \checkmark &                                       &                                \\
   115 & G325.93+00.24 &    12.3 &   N &  1.28 (0.08) &          -- &    2a & \checkmark & \checkmark &  \ding{53} & \checkmark &  \ding{53} &                                       &                 PN               \\
   116 & G326.51+00.03 &    39.6 &  Y? &   9.63 (1.7) & 0.79 (0.31) &   1a? &      $\sim$&      $\sim$&  \ding{53} &  \ding{53} &  \ding{53} &                TYC 8700$-$1257$-$1 (Star) &                                \\
   117 & G328.01+00.52 &    23.9 &   N &  3.47 (0.32) &          -- &    2b & \checkmark & \checkmark &  \ding{53} &          ? &  \ding{53} &                                       &                                \\
   118 & G330.20$-$00.39 &    27.3 &   Y &   1.23 (0.2) & 0.24 (0.08) &    1b & \checkmark &      $\sim$&  \ding{53} &  \ding{53} &  \ding{53} &                                       &                                \\
   119 & G334.95$-$00.08 &    32.7 &   N &  37.3 (1.94) &          -- &    3c & \checkmark &      $\sim$&  \ding{53} &  \ding{53} &          ? &                                       &                   PN             \\
   120 & G335.36+01.04 &    45.9 &   N &  3.23 (0.23) &          -- &    2b & \checkmark & \checkmark &  \ding{53} &        OUT &  \ding{53} &                                       &                   ORC \\
   121 & G336.32$-$00.74 &    59.6 &   N &  51.9 (2.64) &          -- &    2a & \checkmark &      $\sim$&  \ding{53} &  \ding{53} &  \ding{53} &                                       &                             PN \\
   122 & G336.40+00.02 &    19.9 &   N &  52.0 (3.71) &          -- &    2b & \checkmark & \checkmark & \checkmark & \checkmark & \checkmark &                                       &                   Massive star \\
   123 & G336.54+00.60 &    20.3 &   N &  2.92 (0.24) &          -- &    3c & \checkmark &      $\sim$&  \ding{53} &          ? & \checkmark &                                       &                                \\
   124 & G337.10+00.00 &    12.1 &   N &   9.45 (1.4) &          -- &    2a &  \ding{53} & \checkmark &  \ding{53} & \checkmark &          ? &                                       &                     PN           \\
   125 & G338.11$-$00.18 &    21.6 &   N &  38.2 (4.54) &          -- &    2a & \checkmark &      $\sim$& \checkmark & \checkmark & \checkmark &                                       &                     H\textsc{ii} region \\
   126 & G338.17$-$00.11 &    17.3 &   N &  32.9 (2.24) &          -- &    2c & \checkmark & \checkmark & \checkmark & \checkmark & \checkmark &                                       &                                \\
   127 & G339.60+00.44 &    21.3 &   N &  2.31 (0.19) &          -- &    3a &      $\sim$& \checkmark &  \ding{53} &  \ding{53} &  \ding{53} &                                       &                                \\
   128 & G340.59$-$00.72 &    19.2 &  Y? &  5.19 (0.41) &          -- &   1b? &      $\sim$& \checkmark &  \ding{53} &  \ding{53} &  \ding{53} &                                       &                             PN \\
   129 & G340.60$-$00.45 &    15.8 &   N &  0.85 (0.09) &          -- &    3b &      $\sim$&      $\sim$& \checkmark & \checkmark & \checkmark & [SPK2012] MWP1G340603$-$004538 (Bubble) & H\textsc{ii} region / PN / Massive star \\
   130 & G341.00+00.92 &    14.3 &   N &  0.64 (0.06) &          -- &    3b & \checkmark & \checkmark &  \ding{53} &          ? &  \ding{53} &                                       &                                \\
   131 & G341.77$-$00.65 &    19.4 &  Y? &  5.81 (0.32) &          -- &   1b? &  \ding{53} &      $\sim$&          ? &          ? & \checkmark &                                       &                                \\
   132 & G342.76$-$00.33 &    14.6 &   N &  4.16 (0.23) &          -- &    2c & \checkmark & \checkmark &  \ding{53} &          ? &  \ding{53} &        2MASS J16581431$-$4319229 (YSOc) &                                \\
   133 & G343.43+00.19 &    29.7 &   N &  14.7 (0.77) &          -- &    3a &      $\sim$& \checkmark &  \ding{53} & \checkmark &  \ding{53} &                                       &                             PN \\
   134 & G345.23$-$00.44 &    15.6 &   N &  2.19 (0.13) &          -- &    2b &  \ding{53} &  \ding{53} &          ? &          ? &  \ding{53} &                                       &                             PN   \\
   135 & G349.34+00.17 &    11.3 &   N &  1.77 (0.14) &          -- &    3a & \checkmark &      $\sim$&  \ding{53} &  \ding{53} &  \ding{53} &                                       &                             PN \\
   136 & G352.96$-$00.26 &    30.5 &   N &  3.98 (0.34) &          -- &    2a & \checkmark &      $\sim$&  \ding{53} &  \ding{53} &  \ding{53} &                                       &                                \\
   137 & G353.28+00.29 &    18.0 &   N &  11.6 (0.84) &          -- &    2c & \checkmark & \checkmark & \checkmark & \checkmark & \checkmark &                                       &                                \\
   138 & G353.31$-$00.63 &    29.8 &   N &   3.8 (0.25) &          -- &    3b & \checkmark &      $\sim$&  \ding{53} &          ? &  \ding{53} &                                       &                             PN \\
   139 & G355.05$-$01.30 &    28.6 &   N &  2.29 (0.21) &          -- &    3b & \checkmark &      $\sim$&  \ding{53} &  \ding{53} &        OUT &                                       &                             PN \\
   140 & G356.59+00.05 &    55.3 &   Y &  39.2 (2.07) & 0.12 (0.16) &    1a & \checkmark &      $\sim$&  \ding{53} & \checkmark &          ? &     2MASS J17370394$-$3147349 (NearIR ) &                   Massive star \\
   141 & G358.14+00.61 &    15.2 &   N &   1.12 (8.1) &          -- &    3c &      $\sim$& \checkmark &  \ding{53} &  \ding{53} &  \ding{53} &                                       &                    \\
   142 & G358.41+00.11 &    30.0 &   Y &  22.2 (1.24) &  0.25 (0.04) &    1b &  \ding{53} &      $\sim$&  \ding{53} &  \ding{53} &  \ding{53} &                                       &                                \\
   143 & G358.66$-$00.12 &     7.1 &   N &  2.37 (0.14) &          -- &    2a &      $\sim$&      $\sim$&  \ding{53} & \checkmark &  \ding{53} &                                       &                       PN         \\
   144 & G358.67$-$00.26 &     6.5 &   N &  0.76 (0.09) &          -- &    3a & \checkmark &      $\sim$&  \ding{53} & \checkmark &  \ding{53} &                                       &                                \\
   145 & G358.71$-$00.12 &     6.8 &   N &   2.33 (0.2) &          -- &    2a &  \ding{53} &      $\sim$&  \ding{53} &          ? &  \ding{53} &                                       &                                \\
   146 & G359.09$-$01.21 &    10.6 &   N &  1.08 (0.12) &          -- &    3c &  \ding{53} &      $\sim$&  \ding{53} & \checkmark &          ? &                                       &                                \\
   147 & G359.11$-$00.47 &     8.2 &   N &   2.5 (0.14) &          -- &    2c & \checkmark &      $\sim$&  \ding{53} & \checkmark & \checkmark &                                       &                   PN/Massive star \\
   148 & G359.18$-$00.11 &     8.0 &   N &  0.95 (0.11) &          -- &    2b &  \ding{53} & \checkmark &  \ding{53} &  \ding{53} &  \ding{53} &                                       &                                \\
   \midrule
\end{tabular}
\end{table}
\end{landscape}

\begin{landscape}
\begin{table}
\contcaption{}
\footnotesize
\setlength{\tabcolsep}{5pt}
\begin{tabular}{llrclllc@{\hspace{5pt}}cc@{\hspace{5pt}}c@{\hspace{5pt}}cll}
\toprule
 ID & GName & Radius & PS & $S_\mathrm{ring}$ & $S_\mathrm{PS}$ & Type & \multicolumn{2}{|c|}{Point source} & \multicolumn{3}{|c|}{Extended} & SIMBAD & Classification\\
\cline{8-10} \cline{11-12}
\\[-0.4ex]
 &       & (arcsec) & & (mJy) & (mJy) & & 2MASS & GLIMPSE & 8 $\mu$m &  24 $\mu$m & 70 $\mu$m &  \\
\midrule
149 & G359.25$-$00.34 &     6.8 &   N &  1.54 (0.12) &          -- &    2b &  \ding{53} &      $\sim$&  \ding{53} &  \ding{53} &  \ding{53} &                                       &                         PN       \\
   150 & G359.28+00.07 &    12.2 &   N &  1.67 (0.15) &          -- &    3a &      $\sim$&      $\sim$&  \ding{53} &          ? &  \ding{53} &                                       &                   Massive star \\
   151 & G359.36$-$00.00 &     6.6 &   N &  12.1 (0.66) &          -- &    2b &  \ding{53} &      $\sim$&  \ding{53} &  \ding{53} &  \ding{53} &     [YHA2009] 17 (Infrared)                                  &     PN                           \\
   152 & G359.41+00.08 &    10.4 &   Y? &  3.29 (0.21) &          -- &    2b &  \ding{53} &      $\sim$&  \ding{53} &  \ding{53} &  \ding{53} &    SSTGC 255492 (YSOc)                                   &                  H\textsc{ii} region \\
   153 & G359.42+00.30 &     6.7 &   N &  0.81 (0.07) &          -- &    2b & \checkmark &  \ding{53} &  \ding{53} & \checkmark &          ? &        CXOGCS J174301.8$-$291550 (X-ray)                               &           PN                     \\
   154 & G359.55$-$00.29 &     9.9 &   N &  2.71 (0.16) &          -- &    2b &  \ding{53} &      $\sim$&  \ding{53} & \checkmark &  \ding{53} &                                       &                                \\
   155 & G359.58$-$00.06 &     9.0 &   N &  3.16 (0.25) &          -- &    2b &      $\sim$&      $\sim$&  \ding{53} &  \ding{53} &  \ding{53} &                                       &                                \\
   156 & G359.59+00.25 &     8.8 &   N &  1.75 (0.12) &          -- &    2c & \checkmark &      $\sim$&  \ding{53} & \checkmark & \checkmark &        2MASS J17433952$-$2909054 (LPVc) &                   Massive star \\
   157 & G359.62$-$00.13 &     7.9 &   N &  1.84 (0.12) &          -- &    2b &      $\sim$&      $\sim$&  \ding{53} &  \ding{53} &  \ding{53} &                                       &               PN                 \\
   158 & G359.69+00.30 &    11.0 &   Y? &  1.12 (0.12) &          -- &    1b & \checkmark &      $\sim$&  \ding{53} &  \ding{53} &  \ding{53} &                                       &                                \\
   159 & G359.74+00.12 &     6.5 &   N &   1.65 (0.1) &          -- &    2c &  \ding{53} &      $\sim$&  \ding{53} & \checkmark &  \ding{53} &                                       &               PN                 \\
   160 & G359.75$-$00.06 &     9.6 &   N &  16.4 (0.87) &          -- &    2b &      $\sim$&      $\sim$&  \ding{53} & \checkmark & \checkmark &      ISOGAL-P J174518.1$-$291051 (YSOc) &   H\textsc{ii} region / Massive star \\
   161 & G359.81$-$00.02 &     6.7 &   N &  1.03 (0.23) &          -- &    2c &  \ding{53} &      $\sim$&  \ding{53} &  \ding{53} &  \ding{53} &                                       &                   PN             \\
   162 & G359.83+00.06 &     9.3 &   N &  4.14 (0.27) &          -- &    2b &      $\sim$&      $\sim$&  \ding{53} & \checkmark &  \ding{53} &                                       &                   Massive star \\
   163 & G359.89$-$00.52 &    11.1 &   N &   1.5 (0.13) &          -- &    3c &      $\sim$&      $\sim$&  \ding{53} &  \ding{53} &  \ding{53} &                                       &                                \\
   164 & G359.94$-$00.41 &     9.7 &   N &  2.27 (0.16) &          -- &    2b &      $\sim$&      $\sim$&  \ding{53} & \checkmark &  \ding{53} &                                       &                                \\
\bottomrule
\end{tabular}
\end{table}
\end{landscape}

\begin{figure}
    \centering
    \includegraphics[width=\columnwidth]{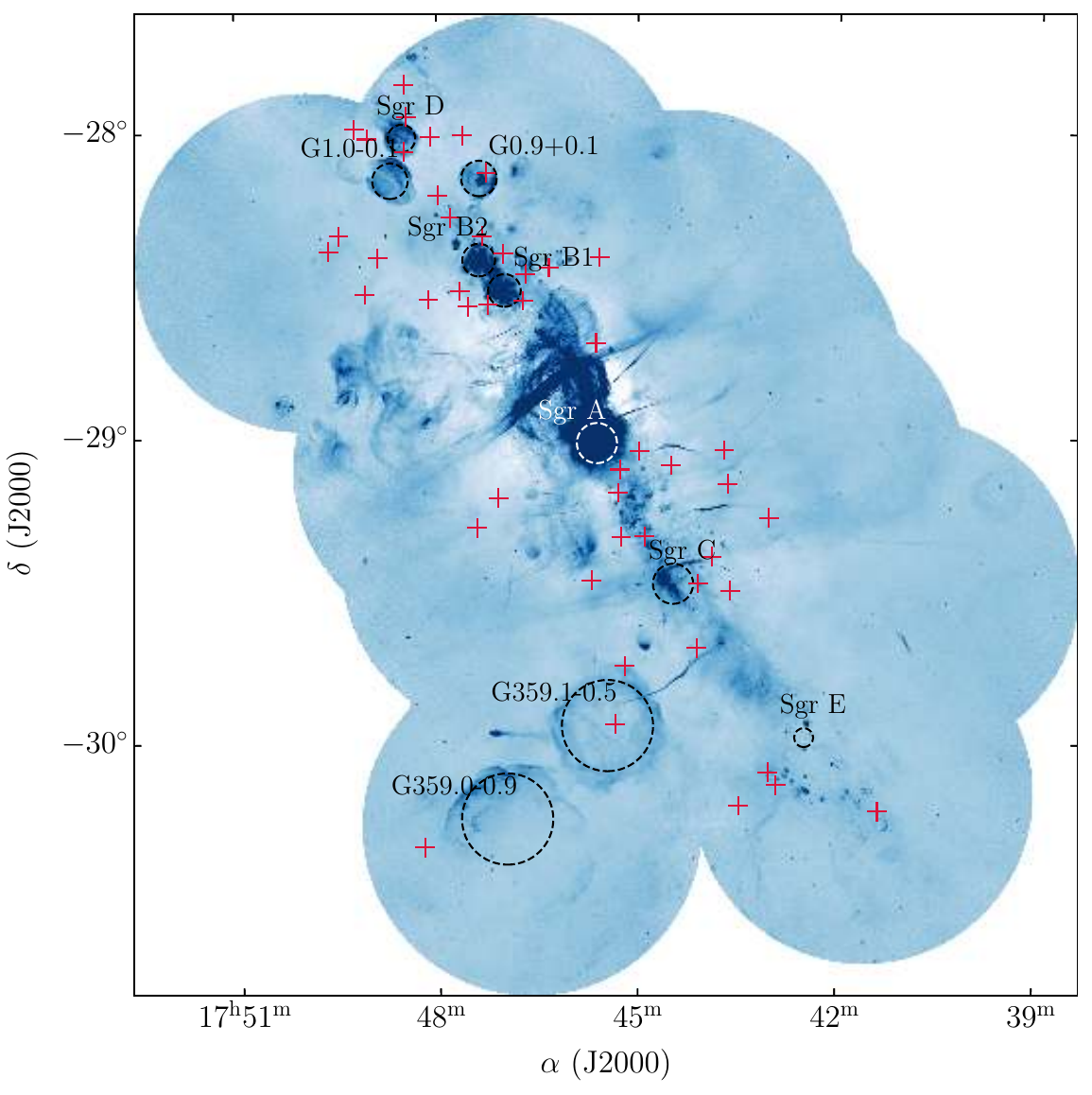}
    \caption{Spatial distribution of radio rings (red crosses) overlaid on the Galactic Centre mosaic \citep{hey2022}. Major star-forming complexes and other prominent radio sources are marked for reference.}
    \label{fig:gal-cen}
\end{figure}

To investigate the impact of resolution-related selection effects, we applied a conservative radius threshold of 10 arcsec, above which both samples can be considered complete (i.e., any ring-like source of $>20$ arcsec diameter would be readily identifiable as a ring in both datasets). There are 12 rings in the Galactic Centre and 114 in the SMGPS that satisfy this criterion, resulting in source densities of 1.8 and 0.23 deg$^{2}$, respectively.
Under these conditions, we applied a Poisson statistical test to assess whether the observed overdensity is significant, assuming the ring density of the SMGPS is representative for the entire Galactic Plane. Under the null hypothesis of no excess, the expected number of rings in the Galactic Centre would be $\lambda=1.5$. The probability of observing 12 or more rings by chance is thus $P(k\geq12 | \lambda =1.5) = 6.8\times10^{-8}$, strongly suggesting that the overabundance has a physical origin, as discussed above. This is supported by the apparent spatial clustering of rings around major star-forming complexes---such as Sgr B1, Sgr B2, Sgr D---visible in the Galactic Centre mosaic, as shown in Fig. \ref{fig:gal-cen}. In addition, the marginal histograms in the top panel of Fig.~\ref{fig:sky-distribution} show a strong concentration of sources around $b=0$ deg, supporting a predominantly Galactic origin for the sample.

\begin{table}
\centering
\caption{Breakdown of rings by morphology type and subtype.}
\label{tab:ring-breakdown}
\begin{tabular}{@{}lccccc@{}}
\toprule
   & \multicolumn{3}{c}{Subtype} & Total & (\%) \\ 
       & a         & b         & c         &       &      \\ \midrule
Type 1 & 9         & 16        & 6         & 31    & 19\% \\
Type 2 & 24        & 51        & 17        & 92    & 56\% \\
Type 3 & 14        & 10         & 17        & 41    & 25\% \\ \midrule
Total  &         &         &         & 164   & 100\% \\ \bottomrule
\end{tabular}
\end{table}

\section{Crossmatch and identification}
\label{sec:crossmatch}

To the best of our knowledge, most---if not all---of the  rings presented in this work are new radio detections, with no prior reports in the literature or obvious matches in existing catalogues of Galactic extended sources (see Sect. \ref{sec:sel-criteria}). In this section, we investigate their multiwavelength footprint to explore plausible astrophysical origins. We note, however, that a definitive classification is beyond the scope of this paper and will require dedicated follow-up observations.

A natural first step in investigating the origin of the rings would be a spectral index analysis. However, despite MeerKAT's L-band providing an effective bandwidth of 856 MHz, in-band spectral indices suffer from exceedingly large uncertainties due to the intrinsic faintness of most rings. Moreover, as discussed in \cite{goe23} and \cite{bor25}, SMGPS spectral indices for extended sources are known to be biased toward negative values. Consequently, we are unable to reliably determine the nature of the radio emission, whether thermal (e.g., free-free) or non-thermal (e.g., synchrotron). Follow-up radio observations at complementary frequencies will be essential to overcome this limitation.

\subsection{SIMBAD search}
We queried the SIMBAD database\footnote{\url{https://simbad.u-strasbg.fr/simbad/}} to identify potential central sources for the rings. Given their location in the crowded Galactic Plane and the high likelihood of spurious associations, we adopted a conservative approach, restricting the search radius to 5 arcsec around the rings' centroids. Within this radius, 28 out of 164 rings have a positional match in SIMBAD ($\sim$17 per cent). The identified counterparts include unclassified sources, galaxies, young stellar object (YSO) candidates, and notably, five \textit{Gaia} long-period variable (LPV) candidates. The ID and type of the matching sources are reported in the second to last column of Table \ref{tab:ring-list}.

\subsection{Multiwavelength footprint}
\label{subsec:multi-footprint}

While the SIMBAD crossmatch provides a useful starting point, it is not exhaustive as it only includes a subset of sources from other catalogs. Therefore, we performed a more comprehensive search for infrared counterparts, including:

\begin{enumerate}
    \item \textit{Central point sources in the near- and mid-infrared}. For rings with detected radio point-like objects, the identification of the central source is trivial. However, rings without a detected radio point source may still host a central object, not visible in the L-band but detectable at shorter wavelengths. To explore this possibility, we performed a search for point sources in the 2MASS (\citealt{skr2006}, II/246/out) and GLIMPSE (surveys I, II and 3D, \citealt{chu2009}, II/293/glimpse) point source catalogues using \texttt{Topcat}\footnote{\url{https://www.star.bris.ac.uk/~mbt/topcat/}}. While 2MASS is an all-sky survey, the sky coverage of GLIMPSE surveys\footnote{GLIMPSE I covers |$l$|=10--65$\degr$ and |$b$|$\leq$1$\degr$; GLIMPSE II covers |$l$|$\leq$10$\degr$, and |$b$|$\leq$1$\degr$ from |$l$|=5--10$\degr$, |$b$|$\leq$1$\fdg$5 for |$l$|=2--5$\degr$ and |$b$|$\leq$2$\degr$ for |$l$|$\leq$2$\degr$; and GLIMPSE 3D extends to |$b$|$\leq$4$\fdg$2 near the Galactic Centre, and up to |$b$|$\leq$3$\degr$ in selected other parts of the Galaxy.} is more limited and does not fully match the footprint of the SMGPS or the Galactic Centre mosaic. Considering the typical size of the rings and the crowdedness of the 2MASS and GLIMPSE fields toward the Galactic Plane, it is almost certain that at least one catalogued point source will fall (in projection) within nearly every ring. We thus restricted the search to a radius of 5 arcsec around the ring centroids and introduced additional quality criteria to mitigate the effects of chance alignments and confusion: for 2MASS, we considered as positive matches only those for which the second nearest catalogued source lies more than 5 arcsec away, consistent with the survey's nominal angular resolution and the catalogue recommendations\footnote{\textit{prox} flag, see \url{https://irsa.ipac.caltech.edu/data/2MASS/docs/releases/allsky/doc/sec2_2a.html\#prox}}; for GLIMPSE, we adopted a similar approach, using the close source flag to identify matches where the second nearest catalogued object is less than 3 arcsec away\footnote{\textit{csf} flag, see \url{https://irsa.ipac.caltech.edu/data/SPITZER/GLIMPSE/gator_docs/GLIMPSE_colDescriptions.html\#csf}}. Matches not meeting these criteria were considered as \lq confused\rq. 
    \item \textit{Extended emission in the mid- and far-infrared}. We visually inspected available \textit{Spitzer} IRAC 8 $\mu$m, MIPSGAL 24 $\mu$m and \textit{Herschel} HiGAL 70 $\mu$m imagery to search for extended emission associated with the radio rings. These wavelengths are particularly informative for studying circumstellar material around evolved stars, as they trace dust populations at different temperatures, from freshly produced ejecta to reprocessed interstellar matter. In our evaluation, we focused on the presence of discrete emission, co-spatial with the radio rings, and  distinguishable from the diffuse background. We considered positive matches even in cases when the infrared emission deviates from a ring-like structure (e.g. diffuse blobs of emission or partial arcs). However, we note that the assessment is incomplete due to the different sky coverage of each survey. The limitation is especially severe for sources at $|b|>1$ deg, for which high-resolution infrared imagery is often unavailable. Furthermore, regions near the Galactic Centre suffer from significant background confusion, which can hamper the identification of faint infrared counterparts.
\end{enumerate}

The results of this search are summarised in Table \ref{tab:ring-list}. The 2MASS and GLIMPSE columns indicate the presence of isolated or confused point sources within 5 arcsec of the centroid, while the 8, 24 and 70 $\mu$m columns denote the presence or absence of extended infrared emission. Our approach identified relatively isolated point sources near the ring centroids for approximately 41 per cent of the rings in 2MASS and 26 per cent in GLIMPSE, whereas confused central sources were found in 26 per cent of the rings in 2MASS and 49 per cent in GLIMPSE. Only 24 rings---representing 15 per cent of the sample---show an isolated central source detected in both surveys. However, the lack of distance information prevents any strong claims about the physical association between these infrared sources and the rings. On the other hand, roughly 50 per cent of the rings display a clear counterpart in at least one of the three infrared bands. Many others, however, exhibit ambiguous or uncertain associations. Only 14 rings---less than 10 per cent of the sample---show a counterpart detected simultaneously in all three bands.

\subsection{What are the rings?}

\subsubsection{Tracers of star formation}

H\textsc{ii} regions can exhibit circular or shell-like morphologies in the radio domain. They typically show characteristic MIR signatures that facilitate their identification: emission at $8-12$ $\mu$m is generally external to the radio continuum, sometimes tracing polycyclic aromatic hydrocarbons (PAHs) excited in photo-dissociation regions \citep{deh10}, while emission at $>20$ $\mu$m, associated with various populations of warm dust, tends to be approximately co-spatial with the radio \citep{and14}.

Several sources in our sample show ring-like morphologies in the radio, but appear embedded within larger-scale structures evident at MIR and FIR wavelengths. These large-scale structures are sometimes part of known H\textsc{ii} regions and complexes listed in the WISE catalogue of Galactic H\textsc{ii} regions (v.2.3, \citealt{and14}). In these cases, the most plausible explanation is that the observed radio ring-like emission corresponds to limb-brightened edges of the larger H\textsc{ii} region, illuminated by the collective ionising radiation from embedded massive stars. Examples include rings 44 (within H\textsc{ii} region G019.629$-$00.095), 59 (G036.993$-$00.231 and G036.959$-$00.223) and 125 (G338.114$-$00.193). 

Other sources, while not co-spatial with catalogued H\textsc{ii} regions, show a similar infrared signature, with the radio continuum emission typically enclosed by diffuse 8 $\mu$m structures. Five rings (47, 56, 100, 125, 129) are partially or fully coincident with infrared \lq bubbles\rq\, identified in the Milky Way Project \citep{sim12, jay19}. These sources may correspond to unconfirmed H\textsc{ii} regions in star-forming complexes, where the observed radio emission is shaped by variations in the ionising field and the relative distribution of dust and gas. Besides, the sensitivity of the SMGPS has recently enabled the detection of weak radio emission from hundreds of H\textsc{ii} regions previously classified as \lq radio-quiet\rq\, in the literature \citep{bor25}.

Additionally, ten rings are probably associated with sources classified in SIMBAD as young stellar object (YSO) candidates (rings 7, 8, 20, 21, 81, 109, 110, 132, 152 and 160). YSOs across the entire mass spectrum can drive outflows and ionise their surroundings, even when deeply embedded or obscured at optical or IR wavelengths. In some cases, radio observations reveal ring- or shell-like morphologies around them. These structures correspond to ultra-compact H\textsc{ii} regions, with typical diameters $\leq$0.1 pc \citep{woo89}. UC H\textsc{ii} regions form when a massive young star begins to photoionise its natal cloud, creating a thin shell of ionised gas that expands under the stellar wind pressure. Indeed, the shell-like morphology is found to be particularly common among UC H\textsc{ii} regions \citep{dep05}, but we found no matches in catalogues of known UC H\textsc{ii} regions \citep{hu16,kal18, men22}. This suggests that the rings associated with YSO candidates may represent previously unreported UC H\textsc{ii} regions. However, definitive confirmation would require a radio follow-up to constrain the radio spectral indices, which are typically in the range  $-0.1\leq$ $\alpha$ $\leq2$ ($S\propto\nu^\alpha$). 

In four of these cases (7, 8, 20, 21), the YSO candidate is notably off-centre, towards the ring edges.  While this might question a physical association, asymmetries in the ionised shell may arise from a non-uniform density structure of the natal cloud. One source---ring 81-- stands out as particularly intriguing: its central object, identified as a class I YSO candidate (2MASS J08520086-4614178), is bright in radio, visible from the far to the near IR, and weakly detected in X-rays with XMM Newton. However, the radio ring itself appears clumpy and lacks a counterpart at other wavelengths, casting doubt on its classification as a YSO-related H\textsc{ii} region.

\subsubsection{Planetary nebulae}
The number of planetary nebulae (PNe) detected in the Milky Way ($\sim$3000, \citealt{sab14}) is an order of magnitude lower than theoretical population estimates \citep{fre06}. PNe are typically discovered at optical wavelengths via their H$\alpha$ emission from circumstellar ionised gas. However, in much of the Galactic Plane, interstellar extinction may obscure them, preventing their detection.

Radio continuum surveys, peering through the intervening dust, play a crucial role in completing the Galactic PNe census, as many PNe are bright at radio wavelengths, showing flat spectral indices \citep{boj11,ira18}. When resolved, they may appear as low-angular diameter ($\leq$1 arcmin) rings or shells without a central point source, which facilitates their identification \citep{ing2016}. Indeed, Ingallinera et al., (in prep.) have identified over 170 PNe candidates in the SMGPS, suggesting that the survey may be recovering many previously undetected low-latitude PNe. Of these, over 30 are coincident with rings in our sample (priv. comm.), which contains many other sources exhibiting radio morphologies consistent with PNe. In principle, most compact rings with bipolar or asymmetric brightness distributions and lacking central sources are likely to be PNe. Based on this criterium, we have selected a subset of promising \lq morphological\rq\, PNe candidates, as listed in Table \ref{tab:ring-list}. However, only observations at complementary radio frequencies can constrain the spectral indices, and confirm the absence of a central point source---which would instead suggest a massive star wind.

\subsubsection{Evolved massive star candidates}

The mass-loss processes of early-type evolved massive stars can produce circumstellar ionised shells. Consequently, many known Galactic LBVs and WRs appear as ring-shaped sources in the radio \citep{dun02}, particularly at the angular resolutions achieved by MeerKAT \citep{goe23} and other SKA precursors \citep{bra25}.  It is therefore possible that our sample includes mass-loss relics from (unidentified) evolved massive stars. Given the rarity of such objects, each new discovery represents a valuable addition to the census of these short-lived evolutionary phases.

Naively, a central point source within a ring  may suggest the presence of a massive star, tracing the thermal emission from its stellar wind. However, this assumption comes with two important caveats. First, without reliable spectral index information we can only speculate on the nature of the point sources, and very different processes (e.g., synchrotron) might explain the observed emission. Second, the absence of a central point source is by no means conclusive, as it might simply reflect sensitivity limitations. The theoretical radio spectrum from a stellar wind scales with $(\dot M / v_\infty)^{4/3}$ and $d^{-2}$ ($\dot M$ being the mass-loss rate, $v_\infty$ the terminal wind velocity, and $d$ the distance, \citealt{pan75}). This implies that, at a given frequency and distance, WRs winds are generally harder to detect than those of LBVs, due to their higher wind velocities and lower mass loss rates.

Another constraint arises from the fact that radio shells associated with LBV and WR stars almost invariably have counterparts at mid- or far-infrared wavelengths, due to the presence of dust formed in their winds and eruptions \citep{koc11}. While this offers an additional means to identify mass-loss relics, the detection of infrared counterparts is not always feasible due to confusion, sensitivity limitations, or lack of sky coverage, as discussed in Sect. \ref{subsec:multi-footprint}.

Five rings in our sample (42, 57, 87, 111 and 156) have a \textit{Gaia} long-period variable (LPV) candidate star projected within 5 arcsec of their geometric centroid. While chance alignment cannot be discarded, such a positional coincidence is suggestive, as long-term variability is one of the defining characteristics of LBV stars. Given that several candidate LBVs in the literature are likewise classified as LPVs in SIMBAD, these sources emerge as potential LBV candidates.

This correspondence prompted us to extend our search to identify other variable stars (not necessarily LPVs) possibly related to the radio rings. We used the catalogue of \textit{Gaia} variables by \cite{mai23}, a curated selection of variable sources in Gaia DR3 brighter than G=17, and the  VISTA Variables in the Via Lactea catalogue DR 4.2 by \cite{sai12}, restricting the search to stellar sources with VARFLAG=1. Using a crossmatch radius of 5 arcsec, we identified 30 initial matches, which were reduced to 22 after excluding those lacking a clear infrared counterpart to the radio ring at 8, 24 or 70 $\mu$m. The resulting associations are presented in Table \ref{tab:variable-stars}. 
 
\begin{figure}
    \centering
    \includegraphics[width=\columnwidth]{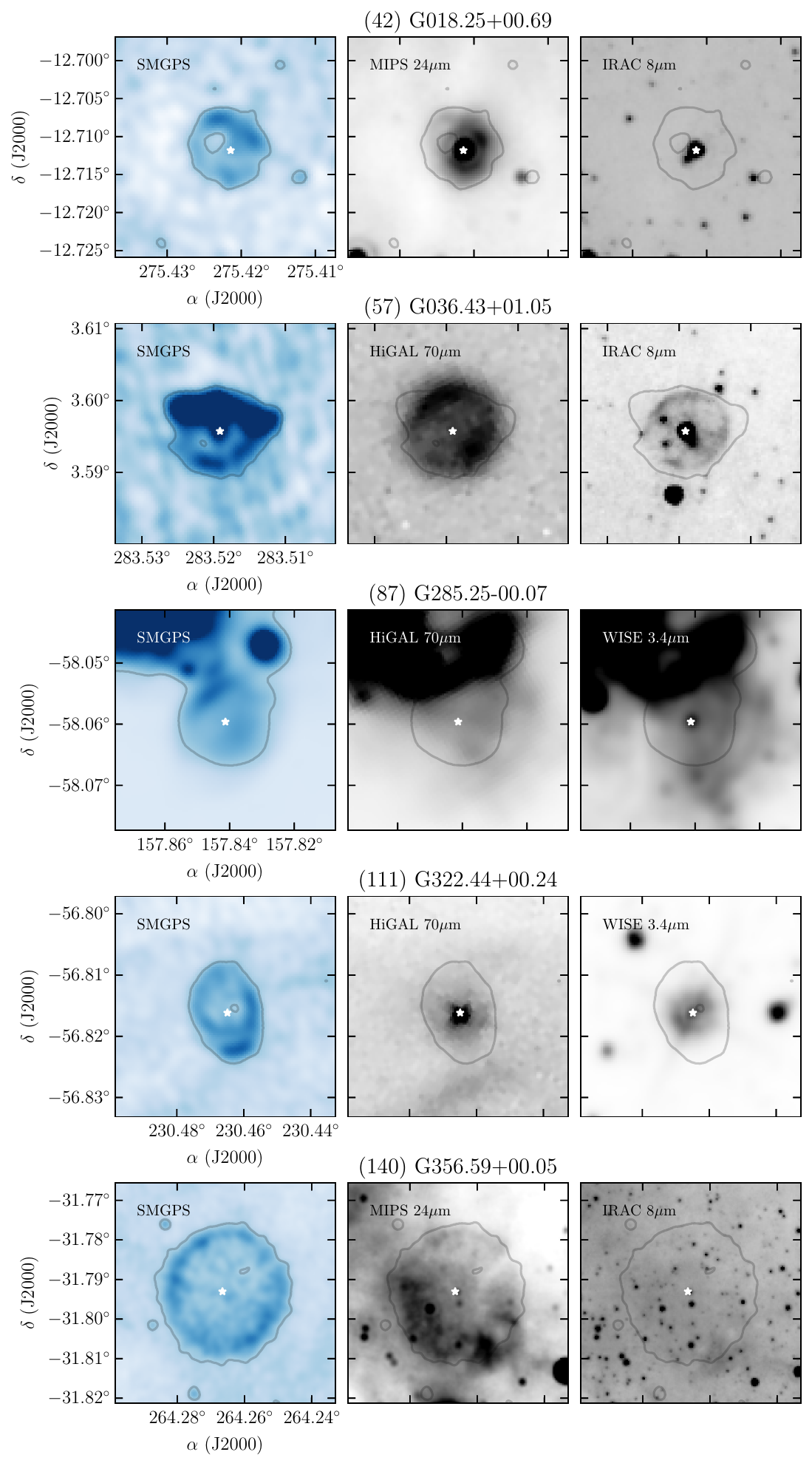}
    \caption{Rings potentially associated with evolved massive stars. The MeerKAT L-band radio continuum is shown in blue colours, while infrared counterparts at different wavelengths (indicated in the top-left corner of each panel) are shown in greyscale. The contours have been chosen for each source to highlight the outline of the radio ring. The position of the coincident variable star is indicated as a white star.}
    \label{fig:candidate-massive}
\end{figure}

Although the relationship between the variable stars and the rings remains speculative, some of the proposed candidates appear especially promising due to their clear infrared signatures, as illustrated in Fig. \ref{fig:candidate-massive}:
 
 \begin{itemize}
 \item Ring 42 shows no clear central radio point source and displays an asymmetric morphology, with a brighter elongated arc towards the northeast. This arc is also visible, though fainter, at 24 $\mu$m, indicating the presence of warm circumstellar dust. At 8 $\mu$m, only the central star is visible, with no signs of extended emission.
 \item Ring 57 presents a well-defined radio shell-like morphology, with a bright central point source. An overlapping elongated feature on the northern edge, resembling a \lq hat\rq, is likely a background radio galaxy, based on its morphology and non-detection in other bands. The shell is prominent at 70 $\mu$m, showing hints of bipolarity, and visible down to 8 $\mu$m, suggesting the presence of warm dust and possibly PAH emission, as observed in other LBVs \citep{uma10}.
 \item Ring 87 has an obvious shell-like morphology  in SMGPS, but its proximity to the bright H\textsc{ii} region G285.260$-$00.051 complicates the search for infrared counterparts.
 \item Ring 111 exhibits an elongated radio morphology in the north-south direction, with a double arc towards the south. While no radio point source is detected, images at 70 and 3.4 $\mu$m reveal an extended dust component slightly internal to the radio shell.
 \item Ring 140 appears as a thin, well-defined circular shell with internal substructure and a faint central source. The shell is clearly detected at 24 $\mu$m, with enhanced brightness along the southeastern edge.
\end{itemize}

At the distances estimated for the coincident stars (see Table \ref{tab:variable-stars}), the radio rings would span physical diameters of approximately $\sim$1 pc, consistent with the typical size of LBV or WR shells. While the multiwavelength features and the presence of likely associated LPVs are encouraging indicators, confirming the definitive nature of these sources will require follow-up observations, including spectroscopy of the central stars.

\begin{table*}
\caption{Rings with co-spatial infrared extended emission and a variable star less than 5 arcsec from the geometric centre. Distances derived from Gaia DR3 parallaxes (error $<30$ per cent and RUWE$<$1.4). Variability references: (1) \protect\cite{gai23}, (2) \protect\cite{mai23}, (3) \protect\cite{sai12}.}
\label{tab:variable-stars}
\begin{tabular}{@{}lllllll@{}}
\toprule
ID & Star                     & $\alpha$         & $\delta$         & Offset & dist.   & Var. Ref. 
\\ 

    &   & (J2000.0) & (J2000.0) & (arcsec) & (kpc) & \\ \midrule
3      & VVV J174643.18$-$283301.84 & 266.679933 & --28.5505110 & 4.4     &      & (3)        \\
12     & Gaia 4057604730105152768 & 267.270273 & --28.5310363 & 3.2     & 2.99 & (2)           \\
15     & VVV J174342.51$-$290230.29 & 265.927122 & --29.0417480 & 1.8     &      & (3)        \\
17     & Gaia 4057611980009974528 & 267.404102 & --28.3896059 & 3.5     & 1.82 & (2)           \\
18     & VVV J174927.89$-$282019.80 & 267.366188 & --28.3388340 & 2.5     &  1.88    & (3)        \\
39     & Gaia 4095676316647302784 & 273.256060 & --18.3080709 & 0.9    & 2.44 & (2) \\
42     & Gaia 4152821543734129536 & 275.421415 & --12.7118523 & 3.2    & 1.54 & (1)           \\
57     & Gaia 4281010885673472256 & 283.519103 & +3.5957213   & 3.5    &      & (1)           \\
87     & Gaia 5351509253451574912 & 157.841258 & --58.0596080 & 1.5    &      & (1)           \\
91     & Gaia 5334372784929698688 & 170.796886 & --61.2264510 & 4.9    & 4.04 & (2)           \\
107     & Gaia 5879230652495048448 & 220.168737 & --58.8344445 & 4.0      & 5.15 & (2)           \\
109    & Gaia 5880016047369386880 & 233.821098 & --56.9449720 & 1.6    & 2.05 & (2)           \\
111    & Gaia 5883324958897499264 & 230.464834 & --56.8161847 & 4.7    & 1.97 & (1)           \\
112    & Gaia 5882675766020323712 & 233.821098 & --56.9449720 & 4.9    & 3.22 & (2)           \\
122    & VVV J163256.78$-$475443.39 & 248.236580 & --47.9120520 & 4.8    &      & (3)        \\
129    & VVV J165108.86$-$450432.56 & 252.786935 & --45.0757120 & 2.6    &      & (3)        \\
140    & VVV J173703.97$-$314734.96 & 264.266543 & --31.7930450 & 3.0    &      & (3)        \\
147    & Gaia 4056844482226725760 & 266.336083 & --29.9405763 & 4.0    & 2.38 & (2)           \\
150    & Gaia 4057068786897090816 & 265.903498 & --29.5021432 & 3.7    & 1.27 & (2)             \\
156    & Gaia 4057147986090205952 & 265.914690 & --29.1515516 & 4.2    &      & (1)           \\
160    & Gaia 4057092014079810432 & 266.325137 & --29.1802397 & 4.8    & 0.83 & (2)           \\
162    & VVV J174459.04$-$290238.72 & 266.245989 & --29.0440890 & 4.5    &      & (3)        \\ \bottomrule
\end{tabular}
\end{table*}

\subsubsection{Supernova Remnants}
\label{sec:SNRs}
Existing catalogues of Galactic SNRs are naturally biased toward large-angular-scale sources due to selection effects in past wide-area radio surveys. SNRs with angular sizes of $\sim$1 arcmin are indeed rare, with fewer than ten confirmed cases and about a dozen candidates \citep[see][their Fig.9]{and25}. Such compact remnants may either be very young or lie at large distances. Given a Galactic supernova rate of one every $40\pm10$ yr$^{-1}$ \citep{tam94}, only $\sim$25 SNRs younger than 1000 yr are expected in the Milky Way \citep{ran21}, in line with current detection statistics. In this context, higher-resolution surveys like the SMGPS may increase the number of compact SNRs.

Recently discovered young SNRs, such as G18.760$-$0.072, G31.299$-$0.493 \citep{ran21} and G329.9$-$0.5 (\textit{Perun}, \citealt{sme24}) show shell-like radio morphologies and weak or absent MIR emission (except for \textit{Perun}, bright at 24 $\mu$m). Several sources in our sample exhibit similar characteristics, such as rings 30, 40 and 120, shown in Fig. \ref{fig:candidate-ORCs}. Despite this resemblance, classifying them as SNRs is problematic. On one hand, their low brightness argues against a young SNR scenario, suggesting they might be older and more distant. On the other hand, however, their small angular diameter contradicts the old SNR interpretation: even at $\sim17$ kpc (on the far side of the Solar circle), the rings would have physical radii of less than 5 pc, exceedingly small for an old SNR.

Ultimately, any classification remains speculative, and only spectral index measurements can truly shed light on the nature of these rings. For reference, Galactic shell-like SNRs typically have spectral indices of $-0.51\pm0.01$ \citep{dub15,ran23}.

Another possibility is that some rings trace pulsar wind nebulae (PWNe). This appears to be the case for ring 82, whose bright central object is identified in SIMBAD as pulsar PSR J0855-4644. The ring appears coincident with the radio nebula reported by \cite{mai18} using uGMRT observations. It lies in the Vela region, a complex area with several overlapping SNRs, complicating the identification of its \lq parent\rq\, remnant. We searched for additional associations in the Green's catalogue of Galactic SNRs and catalogues of candidate SNRs from THOR, GLEAM and SMGPS. Eleven additional rings lie within the boundaries of known or candidate SNRs, namely 27, 31, 43, 48, 56, 71, 130, 131, 138, 140, and 142.  Upon inspection, all appear to be chance alignments, less convincing than the case of ring 82; we find no compelling evidence to suggest a physical association. Similarly, a complementary search in the ATNF pulsar database\footnote{\url{https://www.atnf.csiro.au/research/pulsar/psrcat/}, v2.6.3.} \citep{man05}, yielded no further positional coincidences (apart from PSR J0855-4644).

\subsubsection{Nova Remnants}

Cataclysmic variables (CVs), composed of a white dwarf (WD) accreting material from a donor companion via Roche lobe overflow, undergo sporadic thermonuclear outbursts, known as novae, that result in the ejection of material and the formation of an expanding shell. These shells may remain ionised for extended periods due to UV radiation from the central stars. It is thus conceivable that some of the radio rings in our sample trace old relics from past nova outbursts. This hypothesis was briefly considered for the Kýklos ring in \cite{bor24}, although ultimately discarded.

Nova shells are most commonly detected in the optical, but their overall number remains low. Many H$\alpha$-based searches around CVs have yielded very few detections \citep{sah15,sah22}, suggesting a combination of high recurrence times, intrinsic fading and high extinction. The latter is particularly problematic near the Galactic Plane, where interstellar dust can completely obscure faint nova remnants. In such cases, the radio continuum window would be the only means to uncover them. 

Although a crossmatch with the optical catalogue of nova remnants by \cite{san25} yielded no matches (as most remnants in the catalogue are located at high Galactic latitudes, with only 10 falling within the SMGPS  coverage), we note that several type 1 radio rings in our sample display morphologies strikingly similar to various nova remnants in the catalogue, featuring bright central sources surrounded by diffuse ring-shaped emission. Most nova remnants in \cite{san25}, however, are compact ($r<$0.05 pc) and correspond to recent eruptions, typically younger than $\sim$60 yr. In this sense, the majority of resolved radio images reported in the literature correspond to very young shells, observed within weeks or months after the outburst (e.g., \citealt{Giroletti2020, Lico2024}). That said, larger shells, on the order of a parsec, have also been found around older novae such as V1363 Cyg, Z Cam, and RX Pup. Unfortunately, resolved radio observations of such extended nova remnants are still virtually absent from the literature.

Within our sample, a particularly intriguing candidate is ring 61, one of the lowest surface brightness objects detected. Notably, a Gaia-identified WD candidate lies just $\sim$14 arcsec from its geometric centre. While a physical association cannot be confirmed based on current data, the spatial coincidence is compelling and merits follow-up observations.

That said, interpreting the radio rings as nova remnants is problematic. First, the radio imaging of old nova remnants is challenging as the shells dim as they expand; second, the amount of ejected material in a single nova event is typically modest, of the order of $\sim$10$^{-5}$ M$_\odot$ \citep{yar05}, further limiting detectability. A potential solution involves the secular accumulation of ejecta from recurrent outbursts, forming a nova \lq supershell\rq, as proposed for the extragalactic recurring nova M31 2008-12a \citep{dar19}.

\subsubsection{Background galaxies}
\label{subsec:backgalaxies}

Three sources in the sample are identified as spiral galaxies in SIMBAD: rings 79 (ZOA J084521.622-421845.17), 80 (ZOA J085828.676-451630.99) and 83 (HIZOA J0907-48). In the SMGPS radio images, these galaxies appear ring-like, with the first two featuring a prominent central object. However, infrared imagery reveals a more intricate structure, with spiral arms clearly visible, particularly at 8 $\mu$m. 

These galaxies are located in the so-called \lq Zone of Avoidance\rq\, (ZOA), a region of about $\sim$20 per cent of the sky where the Milky Way dust obscures the extragalactic background in visible light, hindering the detection of optical galaxies. Radio observations are therefore essential to study the ZOA, and SMGPS H\textsc{I} line data \citep{goe23} has already proven effective in identifying hundreds of new galaxy candidates towards the Local Void \citep{kur24}, the Vela Supercluster \citep{raj24} and the Great Attractor \citep{ste24}. We crossmatched our sample with the corresponding catalogues, identifying one match towards the Local Void (64), five towards Vela (79, 80, 83, 84, 86, three of them already in SIMBAD) and three more towards the Great Attractor (99, 102, 106). This brings the number of confirmed galaxies in our sample to nine.

Finally, upon visual examination, we note that rings 94 and 96 exhibit strong 12 $\mu$m emission in WISE W3, similar to rings 79, 80 and 83, and also display hints of spiral morphology at 8 $\mu$m, hence supporting their classification as galaxy candidates. 

\subsubsection{Galaxy cluster lenses}

Strong lensing by galaxy clusters is a powerful probe for the mass distribution in clusters (e.g., \citealt{Acebron2022, Bergamini2023, Acebron2024}).  Giant gravitational arcs by cluster lensing are extremely rare \citep{Meneghetti2013}, but valuable as they can also put strong constraints on the size and compactness of the lens cores (e.g., \citealt{Narayan1996}), and on dark energy (e.g, \citealt{Meneghetti2005, Grillo2024}).  In general, gravitational arcs detected at the radio wavelengths are considerably rarer with respect to the optical wavelengths, as the (background) radio sources are intrinsically rarer (in particular AGN e.g., \citealt{Myers2003, Browne2003, Spingola2019}). 

Since the angular separation scales with the mass of the lens, cluster-scale strong lenses are typically on the order of several arcsec to tens of arcsec. These wide separations provide a unique advantage in resolving lensed images even with interferometers with relatively short baselines. MeerKAT's sensitivity, $uv$-coverage and instantaneous FOV make it well-suited for identifying extended and faint radio emission associated with gravitational arcs, including star-forming galaxies, AGN and HI emission at high-$z$ \citep{Deane2015,Ranchod2022}.

Among the sample of ring-like structures, the most compelling sources to investigate as potential galaxy-cluster lenses are the type 3 objects, given the typical asymmetry of the gravitational potential of galaxy clusters. To possibly confirm some of them as cluster-lenses, high angular resolution multi-band imaging in the optical and near-infrared would be needed, although it can be challenging at low Galactic latitudes due to extinction. Likewise, a spectroscopic follow-up would be critical to determine the redshifts of both the cluster members and the background lensed sources, verifying that the arcs are indeed at higher redshift than the lens.  Finally, the radio image configuration must be explained with a so-called \textsl{lens mass model}. A successful lens model that fits all of the observational constraints would provide strong confirmation that the system is a genuine cluster-scale strong lens, and it will be investigated in a future work.

\subsubsection{Odd Radio Circles}

Odd Radio Circles (ORCs) were first discovered in the EMU ASKAP pilot survey \citep{nor21a}. They appear as well-defined circular radio structures, approximately one arcmin in diameter, with no evident multiwavelength counterparts, negative radio spectral indices ($\alpha<-0.4$), and typically surrounding a distant galaxy ($z\sim$0.5). Following the discovery of the four original ORCs, two additional ORCs have been reported \citep{kor21, nor25}, along with a growing number of ORC-like objects \citep{fil22, kor24, bor24}. Since then, ORCs have become a hot topic in the radio community, with significant efforts devoted to determine their origin. Proposed explanations include shocks from starburst events  \citep{coi24}, supermassive black hole mergers \citep{nor22}, galaxy mergers \citep{dol23}, and remnant radio lobes interacting with external shock fronts \citep{sha24}.

\begin{figure}
    \centering
    \includegraphics[width=\columnwidth]{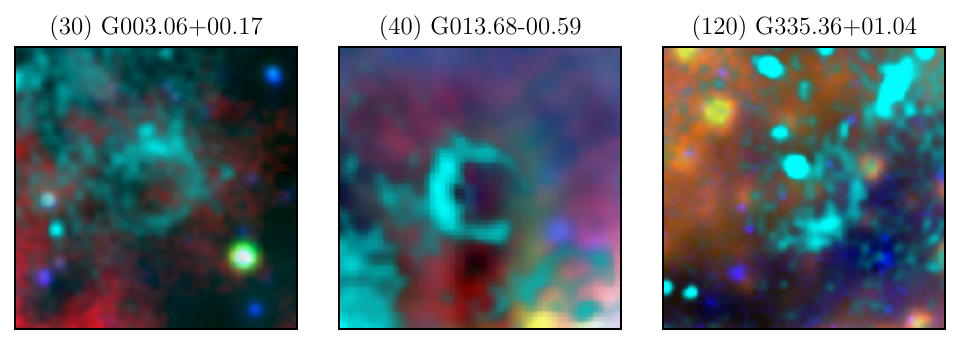}
    \caption{Examples of ring-like structures lacking multiwavelength counterparts. In the absence of further constraints, they may represent ORC candidates or compact SNRs with an unusually low brightness. RGB composite: R=70 $\mu$m, G=24 $\mu$m (except ring 109, where G=WISE 12 $\mu$m due to artifacts), B=8 $\mu$m. MeerKAT L-band radio continuum overlaid in turquoise.}
    \label{fig:candidate-ORCs}
\end{figure}

While most known ORCs and candidates are located at high Galactic latitudes, nothing in principle precludes an ORC from appearing in projection near the Galactic Plane. Indeed, several rings in our sample lack evident multiwavelength counterparts, and, despite the absence of spectral index information, exhibit radio morphologies consistent with those of ORCs---or compact SNRs. These may therefore represent new ORC candidates. Fig. \ref{fig:candidate-ORCs} shows three such examples, namely rings 30, 40 and 120.

\subsubsection{Other possibilities}

The scenarios discussed above represent the most plausible interpretations for the detected radio rings. However, other astrophysical mechanisms may also produce ring-like radio features and warrant consideration.

One such possibility involves high-mass X-ray binary systems (HMXBs). Accreting black holes launch powerful collimated jets capable of inflating shock-driven bubbles in the ISM, extending up to sizes of several parsecs. A well-known example is Cygnus X-1, which is surrounded by a ring-like ionised structure, bright at 1.4 GHz \citep{gal05}. However, crossmatches with catalogues of known Galactic HMXBs \citep{for23} yielded negative results, and due to the limited X-ray coverage across our sample, we are unable to assess this scenario more broadly.

Besides, some of the rings associated with background galaxies may also fit the definition of the so-called starburst radio ring galaxies (RaRiGx), characterised by limb-brightened radio emission outlining the star-forming regions of certain elliptical galaxies, with little or no radio emission arising from the core \citep{gup25}.

\section{Conclusions and future work}
\label{sec:conclusions}

In this work, we have presented a catalogue of 164 radio ring-like structures in the Galactic Plane, discovered with MeerKAT using data from the Galactic Centre mosaic and the SMGPS. These rings, approximately 1 arcmin in size and with flux densities of the order of a few mJy, exhibit a range of features---including central point sources, asymmetries and clumpy substructure---and have been broadly grouped into three morphological classes. The search was performed manually by inspecting the survey tiles, following the selection criteria described in Sect. \ref{sec:sel-criteria}.

Although reliable spectral index information cannot be derived from the in-band MeerKAT data, multiwavelength comparisons with available ancillary data enable a preliminary assessment of the rings' nature. About half the rings exhibit clear counterparts at MIR and FIR wavelengths, others appear embedded in larger-scale structures, and a few are spatially coincident with young stellar objects, known variable stars and unclassified SIMBAD sources. While these associations do not allow for definitive classifications, we explored a range of possible origins, including H\textsc{ii} regions, planetary nebulae, mass-loss relics around evolved massive stars, supernova remnants, nova shells, background galaxies, and even more speculative scenarios, such as ORC candidates or gravitational lensing by galaxy clusters. Tentative classifications have been proposed for 60 per cent of the rings; although these remain provisional given the limitations of the currently available ancillary data. Constraining the true nature of these sources will require dedicated follow-ups, particularly to measure spectral indices and distinguish between thermal and non-thermal emission mechanisms. In this direction, new MeerKAT S-band (1750–3500 MHz) observations of a subsample have already been obtained, and their analysis is underway.

Regardless of their individual classification, the discovery of these rings demonstrates the potential of MeerKAT and, more generally, SKA precursors, to reveal previously unknown, low-surface brightness radio sources, holding great promise for unexpected discoveries in the coming years, as MeerKAT's capabilities continue to expand.

\section*{Data availability}
All SMGPS DR1 data products are available at \url{https://doi.org/10.48479/3wfd-e270}. The Galactic Centre Mosaic data products can be accessed at \url{https://doi.org/10.48479/fyst-hj47}. The complete catalogue of ring-like sources will be made available at <link provided upon acceptance>.

\section*{Acknowledgements}
We thank the anonymous referee for their constructive and insightful comments and suggestions, which helped to strengthen this paper. The MeerKAT telescope is operated by the South African Radio Astronomy Observatory, which is a facility of the National Research Foundation, an agency of the Department of Science and Innovation. This work was supported in part by the Italian Ministry of Foreign Affairs and International Cooperation, grant number ZA23GR03. C.~B. acknowledges financial support from INAF $-$ Ricerca Fondamentale 2024 Mini Grant program (Ob. Fu. 1.05.24.07.02). C.~S. acknowledges financial support from INAF $-$ Ricerca Fondamentale 2024 (Ob.Fu. 1.05.24.07.04) and by the Italian Ministry of University and Research (grant FIS-2023-01611, CUP C53C25000300001). This research has made use of the Spanish
Virtual Observatory (\url{https://svo.cab.inta-csic.es}) project funded by
MCIU/AEI/10.13039/501100011033/ through grant PID2023-146210NB-I00.

\bibliographystyle{mnras}
\bibliography{references} 

\appendix
\section{Notes on individual sources}
\label{app:notes}

\noindent 2. The ring is embedded in a large-scale ISM cavity, clearly visible in the 70 $\mu$m image.

\noindent 3. The ring sits next to one of the bright non-thermal radio filaments of the Galactic Centre \citep{hey2022}. SIMBAD reports a catalogued X-ray source at the edge of the ring, CXOGCS J174642.6$-$283305.

\noindent 4. In the radio, an elongated structure approximately 50 arcsec in size appears tangent to the ring. It is not detected at any other wavelength and may be a background galaxy.

\noindent 7. The YSO candidate ISOGAL-P J174733.0$-$283411 lies within the boundaries of the ring, but is not coincident with the central radio source. It accounts for most of the emission seen at 24 $\mu$m, suggesting a possible chance alignment.

\noindent 8. The peak of 24 $\mu$m emission is correlated with the radio maximum.

\noindent 9. The radio emission is isolated but the emission at 70 $\mu$m forms part of a larger structure.

\noindent 10. The ring lies within a large-scale cavity in the 70 $\mu$m image.

\noindent 15. The bright H\textsc{ii} region [KC97c] G000.9+00.1 lies $\sim$40 arcsec southeast of the source.

\noindent 16. Star [FPF2021] 56 20 1 is located on the southeastern edge of the ring.

\noindent 21. A bright GALEX UV source is located on the southern edge of the ring.

\noindent 22. The source is located next to one of the non-thermal radio filaments of the Galactic Centre.

\noindent 24. The submillimetre source JCMTSF J174914.0$-$275917 is located on the northern edge of the ring.

\noindent 27. Star [FPF2021] 52 5 1 appears to be coincident with the western side of the ring.

\noindent 28. The ring is elongated in the east--west direction. Although no obvious co-spatial MIPS 24 $\mu$m emission is detected, there is a depression towards the ring centroid.

\noindent 30. A weak H$\alpha$ point source is detected in the SuperCOSMOS H$\alpha$ Survey \citep{par05} toward the ring centroid.

\noindent 32. In the 24 $\mu$m image, faint emission is present that may be attributed to the ring.

\noindent 33. Discrete emission at 24$\mu$m is visible towards the ring.

\noindent 34. A \textit{Gaia} LPVc (2MASS J17515821$-$2438216) lies $\sim$10 arcsec from the centroid, near the southwestern edge of the ring, and is most likely unrelated.

\noindent 35. A yellowish PanSTARRS \citep{kai02} source is located near the geometric centre of the ring.

\noindent 36. The field is dominated by the bright source IRAS 17540$-$2434, a \textit{Gaia} LPVc  unrelated to the ring.

\noindent 37. The ring bipolar morphology is clearly resolved at 24 $\mu$m.

\noindent 40. A reddish PanSTARRS source is located near the geometric centre of the ring.

\noindent 42. The \textit{Gaia} LPV candidate 2MASS J18214114$-$1242424 lies at the geometric centre of the ring. The 24 $\mu$m emission closely traces the radio, with a particularly bright arc toward the northeast.

\noindent 43. At 8 and 70 $\mu$m, a bright, dusty arc is visible along the northeastern edge of the ring. The X-ray source XGPS-I J182436$-$115127 lies within the ring.

\noindent 44. In the radio band, the ring appears superimposed on the much larger H\textsc{ii} region G019.629$-$00.095. The southern edge of the ring is particularly prominent in the infrared. A faint H$\alpha$ point source in SHS is coincident with the ring's centroid.

\noindent 45. Very faint 24 $\mu$m emission appears towards the ring.

\noindent 46. The central stellar source 2MASS J18264481$-$1106216 is bright at H$\alpha$ wavelengths.

\noindent 47. The ring, particularly bright toward the south at 70 $\mu$m, is coincident with the infrared bubble [JDP2019] MWP2G0217233+0024908.

\noindent 49. The ring is located near a bright radio clump to the northeast.

\noindent 50. The radio ring is elongated in the East-West direction. The YSO candidate 2MASS J18354814$-$0807235 lies within the ring but offset from the centroid.

\noindent 51. Weak extended emission at 70 $\mu$m, part of a larger structure, may be associated with the radio ring. A weak SHS source is located near the geometric centre, and the sub-millimetric source [ERG2015] 4003 is $\sim$6 arcsec away.

\noindent 52. The faint 70 $\mu$m emission seems to be part of a larger dust structure.

\noindent 53. The central source visible in 2MASS is also detected at H$\alpha$ wavelengths.

\noindent 56. The co-spatial 70 $\mu$m emission is identified as bubble [SPK2012] MWP1G032714+000291.

\noindent 57. A background radio galaxy overlaps the ring on its northern side. The ring is bright at mid- and far-infrared wavelengths. The central point source, classified as a LPVc, is very bright in 2MASS and IRAC.

\noindent 58. In the radio, the ring appears isolated, but the co-spatial infrared emission seems to be part of a much larger dusty structure, the dark infrared nebula DOBASHI 1552.

\noindent 59. The radio ring is elongated in the southeast-northwest direction. The region is bright and filamentary in the infrared, possibly being linked to molecular cloud CHIMPS 3647.

\noindent 60. The ring lies within a large-scale cavity visible at infrared and radio wavelengths.

\noindent 61. The white dwarf candidate \textit{Gaia} DR2 4281278307515970560 is located $\sim$14 arcsec away from the ring centroid.

\noindent 62. There is an optical source near the geometric centre of the ring, Gaia DR3 4281368467478108928.

\noindent 64. The radio point source is much brighter than the diffuse ring. It is detectable down to IRAC wavelengths, with hints of the shell visible at 8 $\mu$m. The galaxy ALFA ZOA J1901+0651 is located $\sim$30 arcsec north.

\noindent 65. The ring is partially filled and notably circular. A slightly elongated central source is tentatively visible.

\noindent 69. The ring is slightly elongated in the east--west direction.

\noindent 70. The bright blue star TYC 1049$-$405$-$1 lies less than 10 arcsec away from the ring's centroid.

\noindent 71. The bipolar radio ring is located adjacent to the bright infrared complex associated with the sub-millimetric source BGPSv2 G046.170$-$00.098.

\noindent 72. At both 24 $\mu$m and in the radio, the ring displays hints of an \lq S-shaped\rq\, jet-like feature, aligned with the approximate direction of the symmetric brightness peaks.

\noindent 74. The giant H\textsc{ii} region G049.484$-$00.391 is adjacent to the ring, with a bright clump partially overlapping it.

\noindent 75. The 2MASS source 19295096+1934205 lies within the ring, $\sim$8 arcsec from the radio point source.

\noindent 76. The westernmost edge of the ring is bright in the infrared, visible from 8 $\mu$m to 70 $\mu$m. The submillimetre source HIGALBM G055.2087$-$00.5265 is located within the ring.

\noindent 77. The ring is partially blended with a diffuse structure to the northeast, that is detected at 8 $\mu$m. Weak emission is present toward the ring, detected also in H$\alpha$, though not perfectly correlated.

\noindent 78. The radio ring is slightly brighter along its northern edge. There is a reddened 2MASS source $\sim$14 arcsec off-centre, faint but still visible in SHS H$\alpha$.

\noindent 79, 80, 83. At IRAC 8 $\mu$m, spiral arms surrounding a bright core are seen, consistent with their SIMBAD classification as ZOA galaxies (see Sect. \ref{subsec:backgalaxies}).

\noindent 81. The bright central source, which outshines the ring, is clearly detectable from 70 $\mu$m to 8 $\mu$m, and coincides with the YSO candidate 2MASS J08520086$-$4614178. It is also coincident with the soft X-ray source 4XMM J085200.8$-$461418.

\noindent 82. The central point source is coincident with a soft X-ray source 4XMM J085536.2-464414, identified as the pulsar PSR J0855-4644.

\noindent 84. The central radio point source is detected at 8 $\mu$m, along with some hints of the extended structure.

\noindent 85. There is no prominent central source, hints of filamentary extended structure are detected at 8 $\mu$m. The AGN candidate HIZOA J0952$-$55A is located within the ring, $\sim$12 arcsec from the centroid.

\noindent 87. The ring is partially blended with the overlapping giant H\textsc{ii} region G285.260$-$00.051, both in the radio and at 70 $\mu$m. The central point source, identified as the \textit{Gaia} LPVc 2MASS J10312191$-$5803345, is barely detected in the optical DSS2 red band.

\noindent 88. A filamentary structure at 8 $\mu$m enshrouds the smooth radio ring. The field is crowded, with multiple 2MASS sources  within the ring.

\noindent 89. A faint H$\alpha$ source is visible in SHS near the geometric centre of the ring.

\noindent 90. The radio morphology is peculiar, depicting a faint ring with two bright knots. Weak, non-catalogued 2MASS emission is visible near the geometric centre.

\noindent 93. The radio ring seems to be embedded in a large-scale infrared structure. The YSO candidate 2MASS J12270006-6305285 lies on the eastern edge of the ring.

\noindent 94. A bright SHS H$\alpha$ source lies about 3 arcsec away from the geometric centre of the ring.

\noindent 95. The ring is elliptical, elongated in the north--south direction.

\noindent 99. The source is positionally coincident with the galaxy GLADE 1350990.

\noindent 100. The ring is bright at mid-infrared wavelengths, and appears associated with bubble [JDP2019] MWP2G3134593$-$0013236.

\noindent 104. The symmetric bright blobs in the radio ring are coincident with two point sources visible at 8 $\mu$m.

\noindent 105. The ring is irregular, with a significantly brighter eastern side. At other wavelengths, the field is dominated by the very bright source IRAS 14249$-$5952, identified as a LPVc and located near the brighter edge.

\noindent 106. The radio ring contains an elongated interior source that does not appear point-like. However, the feature is point-like at 8 and 24 $\mu$m, and corresponds to a highly reddened 2MASS source, identified as galaxy GLADE 1350852.

\noindent 107. The central point-like source is much brighter than the diffuse ring.

\noindent 108. The weak 70 $\mu$m counterpart is diffuse and does not show a ring-like morphology.

\noindent 109. At infrared wavelengths, the ring counterpart appears compact with a tail-like feature extending eastward. It is coincident with YSO candidate 2MASS J15084437$-$5842399.

\noindent 110. The ring lies close to a very bright infrared source, identified as bubble [SPK2012] MWP1G320333$-$003053.

\noindent 111. The radio ring is asymmetric, with a brighter southern edge. It displays compact internal 70 $\mu$m emission, coincident with the position of the LPVc  IRAS 15179$-$5638.

\noindent 112. The radio ring shows a western elongation reminiscent of the structures found by \cite{ing2019} in the evolved massive star candidates SCO J165412$-$410032 and SCO J165433$-$410319.

\noindent 114. The radio ring shows a peculiar morphology, featuring two \lq connected\rq lobes and hints of a central source. Isolated 70 $\mu$m emission is clearly detected towards the ring.

\noindent 116. A point source lies within the radio ring, significantly off-centre, toward the southwestern side. This source may be associated---within uncertainties--- with a bright source visible down to H$\alpha$ wavelengths (2MASS 15462115$-$5437130). The star TYC 8700$-$1257$-$1 is located near the geometric centre of the ring.

\noindent 117. At 24 $\mu$m, two faint, symmetric bands of emission may correspond to the brightest edges of the radio ring.

\noindent 118. The radio ring is asymmetric, almost \lq C-shaped\rq, with an opening towards the southeast.

\noindent 119. Diffuse emission at 70 $\mu$m is seen towards the ring, but with an irregular morphology.

\noindent 121. A bright, off-centre radio point source lies within the ring, though its association is unclear.

\noindent 122. At all wavelengths, the ring appears embedded in a concentric, large-scale structure, the molecular cloud [RC2004] G336.4+0.0-128.5. Within the ring lies the LPVc 2MASS J16325757$-$4754520.

\noindent 123. Weak hints of the radio ring are visible at 24 and 70 $\mu$m.

\noindent 125. The radio ring overlaps with the H\textsc{ii} region G338.114$-$00.193. The co-spatial mid- and far-infrared emission exhibits a different, more filamentary morphology, likely associated with the bubble [SPK2012] MWP1G338120$-$001882.

\noindent 126. The strong bipolarity of the ring is also observed at 24 $\mu$m. The central 2MASS source is visible down to H$\alpha$ wavelengths.

\noindent 127. A bluish 2MASS source, bright in H$\alpha$, is located slightly off-centre.

\noindent 128. The eclipsing binary Gaia DR3 5964190427032327552 is located on the northern side of the ring, though its association is unlikely.

\noindent 129. The radio emission appears confined within a larger infrared ring, identified as bubble [SPK2012] MWP1G340603$-$004538.

\noindent 131. The ring is embedded in a slightly larger diffuse infrared cloud, primarily visible at 24 $\mu$m.

\noindent 132. Central emission, not visible in the radio, is clearly detected at 8 and 24 $\mu$m, likely associated with the YSO candidate 2MASS J16581431$-$4319229.

\noindent 133. The 24 $\mu$m follows a radial profile and is slightly less extended than the radio ring. This emission is associated with the infrared source IRAS 16546$-$4223.

\noindent 134. The brightest side of the radio ring is only tentatively detected at 24 $\mu$m. The bright spectroscopic binary UCAC4 243$-$115623 is located within the ring, $\sim$10 arcsec from the centroid.

\noindent 135. The bright H\textsc{ii} region G349.336+00.147 is located south of the ring. An extremely faint H$\alpha$ source is within the ring.

\noindent 138. There is tentative detection of extremely faint 24 $\mu$m emission associated with the ring.

\noindent 139. The central 2MASS point source is detected in H$\alpha$.

\noindent 140. The radio ring is thin and well-defined, with a weak point source in the centre. The ring is faint but detectable at 24 $\mu$m. The central source, detected down to H$\alpha$ wavelengths, and corresponds with source 2MASS J17370394$-$3147349.

\noindent 142. The radio ring exhibits a complex, clumpy morphology, with a central point source standing out. It is located, in projection, within a large scale filamentary structure. 

\noindent 144. The radio ring is affected by some nearby imaging artefacts, but a weak infrared counterpart clearly stands out in the 24 $\mu$m images.

\noindent 145. The radio ring is located next to the H\textsc{ii} region G358.72$-$00.127.

\noindent 146. The radio morphology appears clumpy, possibly due to its faintness and the noisy background, but the corresponding emission at 24 $\mu$m is smooth and roundish.

\noindent 147. Extended emission associated with the radio ring is detected down to 8 $\mu$m. The Chandra soft X-ray source 2CXO J174520.5$-$295624 is located within the ring.

\noindent 150. The ring shows an open \lq C\rq, morphology and lies between two Galactic Centre filaments.

\noindent 151. The IR source [YHA2009] 17, classified as a YSO candidate, is positionally coincident with the ring. A bright radio point source is adjacent to the ring.

\noindent 152. The ring is brighter towards the eastern side and lies close to a radio filament.

\noindent 153. The X-ray source CXOGCS J174301.8-291550 covers the position of the radio ring.

\noindent 154. The ring shows an open \lq C\rq, morphology in the radio. Weak, diffuse emission within the ring is visible in PanSTARRS.

\noindent 156. The ring is bipolar and slightly elliptical. The LPV candidate 2MASS J17433952$-$2909054 lies on the northeastern edge, but its position relative to the ring makes its association unlikely.

\noindent 158. The ring shows an asymmetric brightness distribution and is coincident with a SHS H$\alpha$ point source.

\noindent 160. The YSO candidate ISOGAL-P J174518.1$-$291051 is located on the brightest edge of the ring.

\noindent 162. The \textit{Gaia} LPV candidate 2MASS J17445907$-$2902378 is located on the brightest edge of the ring.

\noindent 163. The ring has a clumpy morphology with slight ellipticity. A bluish PanSTARRS source lies within the ring.

\noindent 164. The ring morphology is clearly distinguishable at 24 $\mu$m.

\clearpage
\section{Radio continuum images of the rings}
\label{app:images}

Fig. \ref{fig:all-sources-part-1} shows L-band cutouts for all the rings in the catalogue.

\begin{figure*}
    \centering
    \includegraphics[width=0.85\textwidth]{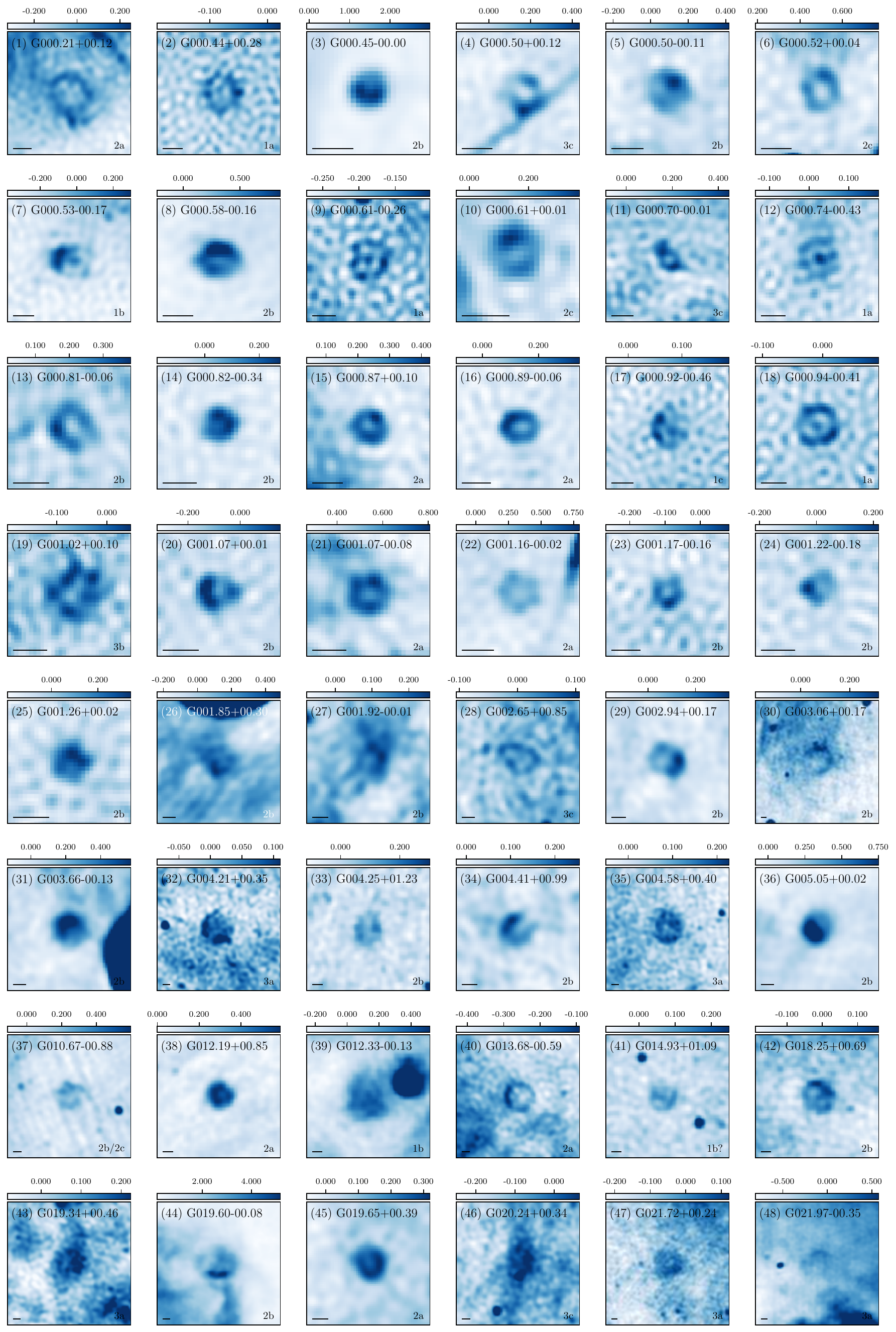}
    \caption{Cutout images of the rings. Each panel is labelled with the source ID and name in the top-left corner. Plots are presented in equatorial coordinates. The morphological type is indicated in the bottom-right corner. The scale bar corresponds to 10 arcsec. The colour bar indicates surface brightness in units of mJy beam$^{-1}$.} \label{fig:all-sources-part-1}
\end{figure*}

\begin{figure*}
  \ContinuedFloat
    \centering
    \includegraphics[width=0.85\textwidth]{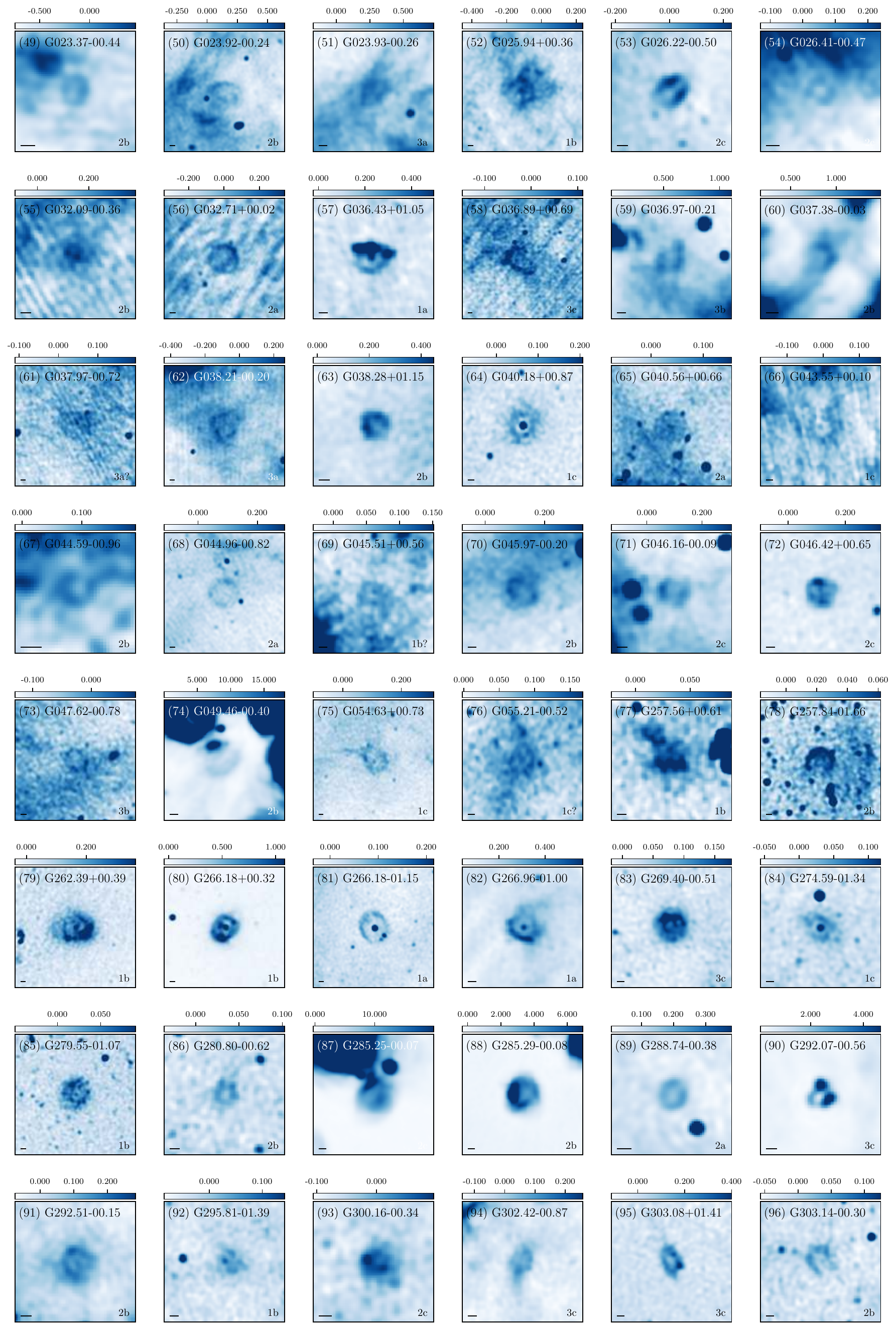}
    \caption{(continued).}
\end{figure*}

\begin{figure*}
  \ContinuedFloat
    \centering
    \includegraphics[width=0.85\textwidth]{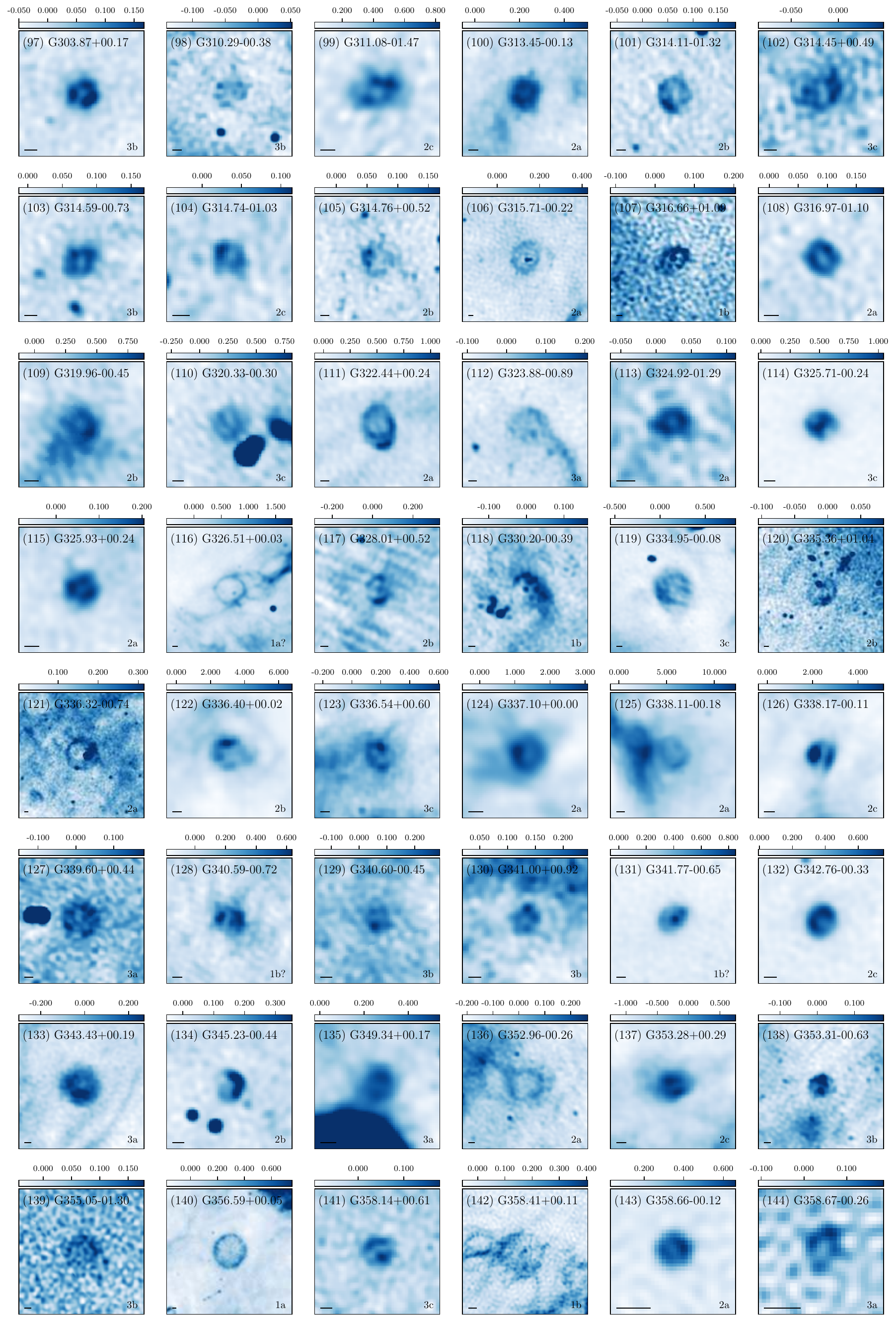}
    \caption{(continued).}
\end{figure*}

\begin{figure*}
  \ContinuedFloat
    \centering
    \includegraphics[width=0.85\textwidth]{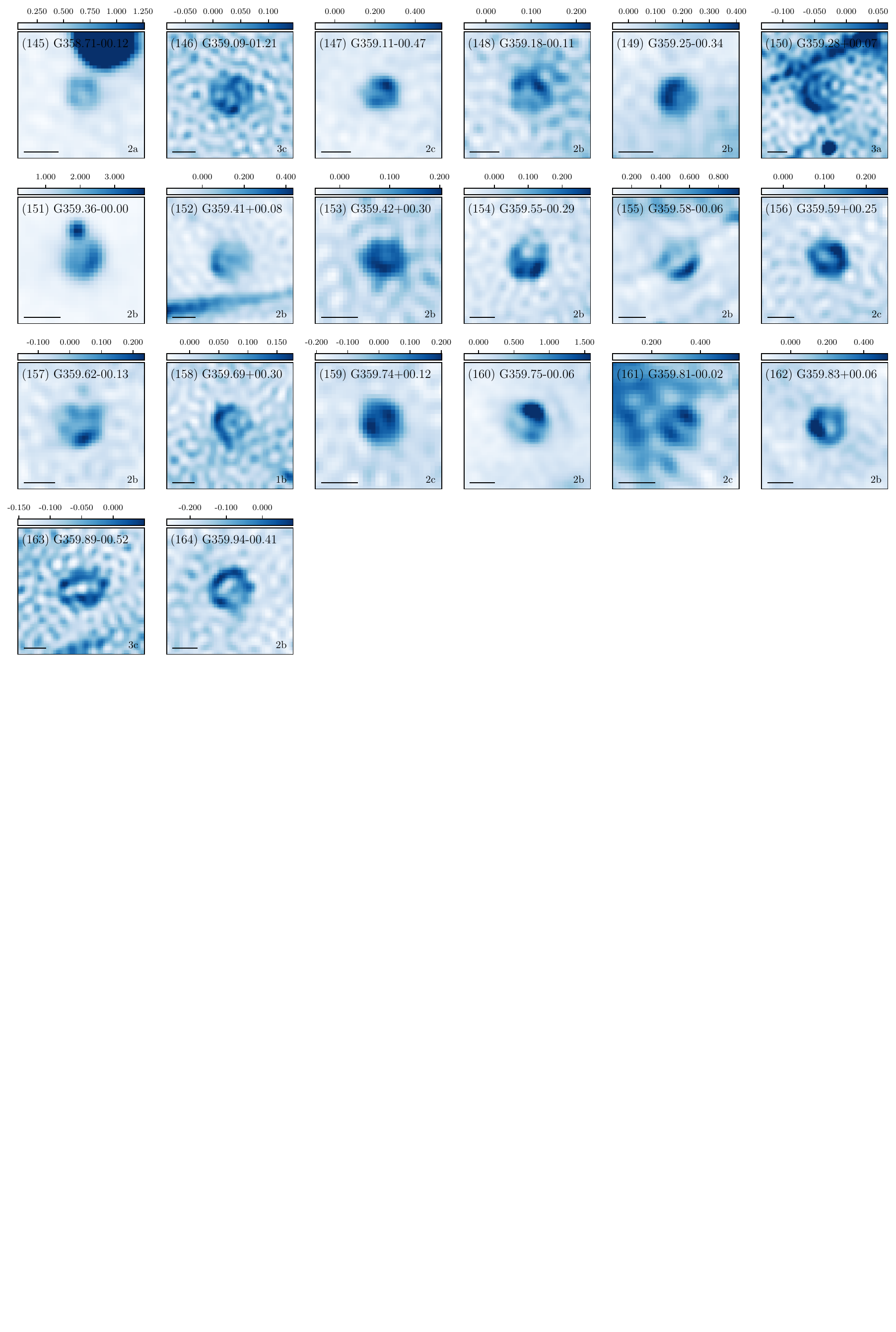}
    \caption{(continued).}
\end{figure*}


\bsp	
\label{lastpage}
\end{document}